\documentclass[acmsmall]{ec20acm}

\newcommand{\squishlist}{\begin{itemize}}
\newcommand{\squishend}{\end{itemize}}

\usepackage{graphicx}
\usepackage{epstopdf}
\usepackage{epsfig}
\usepackage{booktabs} 
\usepackage{amsmath}
\usepackage{amssymb}
\usepackage{amsthm}
\usepackage{amscd}
\usepackage{amsfonts}
\usepackage{graphicx}%
\usepackage{fancyhdr}
\usepackage{color}
\usepackage{algpseudocode}
\usepackage{algorithm}
\usepackage[british]{babel}
\usepackage{enumitem}
\usepackage{bbm}
\usepackage{caption,subcaption}
\usepackage{wrapfig}
\usepackage{multirow}

\def\BState{\State\hskip-\ALG@thistlm}

\makeatletter  
 \renewcommand{\ALG@name}{Mechanism} 
\makeatother   

\newtheorem{corollary}{Corollary}
\newtheorem{assumption}{Assumption}
\newtheorem{remark}{Remark}

\usepackage{hyperref}

\newcommand{\E}{\mathbb E}

\newcommand{\Sc}{R}
\newcommand{\M}{\mathcal M}

\newcommand{\RNum}[1]{\uppercase\expandafter{\romannumeral #1\relax}}
\newcommand{\sgne}[2]{e^{#1}_{#2}}

\newcommand{\pose}[1]{e^{+}_{#1}}

\newcommand{\nege}[1]{e^{-}_{#1}}
\newcommand{\posz}[1]{\alpha_{#1}}
\newcommand{\mat}[1]{\beta_{#1}}
\newcommand{\mathree}[1]{\gamma_{#1}}

\newcommand{\D}{\mathcal{D}}

\renewcommand{\S}{\mathcal{S}}
\newcommand{\za}{z_1}
\newcommand{\zb}{z_2}
\newcommand{\zc}{z_3}

\newcommand{\SSRM}{{SSR mechanism} }
\newcommand{\SSRalpha}{\text{$\text{SSR}_\alpha$} }

\ifodd 0

\newcommand{\com}[1]{\textbf{\color{red}(COMMENT: #1)}} 
\newcommand{\clar}[1]{\textbf{\color{green}(NEED CLARIFICATION: #1)}}
\newcommand{\response}[1]{\textbf{\color{magenta}(RESPONSE: #1)}} 
\newcommand{\jt}[1]{{\color{red}(Juntao: #1)}}
\newcommand{\yc}[1]{{\color{blue}(Yiling: #1)}}

\else

\newcommand{\com}[1]{}
\newcommand{\clar}[1]{}
\newcommand{\response}[1]{}
\newcommand{\jt}[1]{}
\newcommand{\yc}[1]{}
\fi

\usepackage{xcolor}
\usepackage{colortbl}
\definecolor{best}{HTML}{BAFFCD}
\definecolor{secondbest}{HTML}{FFFF66}
\definecolor{issue}{HTML}{FFC8BA}
\definecolor{bad}{HTML}{FFC8BA}

\newcommand{\titlecell}[2]{\setlength{\tabcolsep}{0pt}{\textbf{{\begin{tabular}{#1}#2\end{tabular}}}}}

\AtBeginDocument{%
  \providecommand\BibTeX{{%
    \normalfont B\kern-0.5em{\scshape i\kern-0.25em b}\kern-0.8em\TeX}}}

\copyrightyear{2020} 
\acmYear{2020} 
\setcopyright{acmlicensed}\acmConference[EC '20]{Proceedings of the 21st ACM Conference on Economics and Computation}{July 13--17, 2020}{Virtual Event, Hungary}
\acmBooktitle{Proceedings of the 21st ACM Conference on Economics and Computation (EC '20), July 13--17, 2020, Virtual Event, Hungary}
\acmPrice{15.00}
\acmDOI{10.1145/3391403.3399488}
\acmISBN{978-1-4503-7975-5/20/07} 

\begin{document}

\title{Surrogate Scoring Rules}

\author{Yang Liu}
\authornote{Both authors contributed equally to this research.}
\email{yangliu@ucsc.edu}
\affiliation{%
  \institution{UC Santa Cruz}
  \country{USA}
}
\author{Juntao Wang}
\authornotemark[1]
\email{juntaowang@g.harvard.edu}
\affiliation{%
  \institution{Harvard University}
  \country{USA}
}

\author{Yiling Chen}
\email{yiling@seas.harvard.edu}
\affiliation{%
  \institution{Harvard University}
  \country{USA}
 }

\renewcommand{\shortauthors}{Liu and Wang, et al.}

\fancyhead{} 

\begin{abstract}
Strictly proper scoring rules (SPSR) are incentive compatible for eliciting information about random variables from strategic agents when the principal can reward agents after the realization of the random variables. They also quantify the quality of elicited information, with more accurate predictions receiving higher scores in expectation. In this paper, we extend such scoring rules to settings where a principal elicits private probabilistic beliefs but only has access to agents' reports. We name our solution \emph{Surrogate Scoring Rules} (SSR). SSR build on a bias correction step and an error rate estimation procedure for a reference answer defined using agents' reports. We show that, with a single bit of information about the prior distribution of the random variables, SSR in a multi-task setting recover SPSR in expectation, as if having access to the ground truth. Therefore, a salient feature of SSR is that they quantify the quality of information despite the lack of ground truth, just as SPSR do for the setting \emph{with} ground truth. As a by-product, SSR induce \emph{dominant truthfulness} in reporting. Our method is verified both theoretically and empirically using data collected from real human forecasters.
\end{abstract}

\ccsdesc[500]{Information systems~Incentive schemes}
\ccsdesc[500]{Theory of computation~Quality of equilibria}

\keywords{Strictly proper scoring rules, information elicitation without verification, peer prediction, dominant strategy incentive compatibility, information calibration}

\maketitle

\section{Introduction}\label{sec:intro}

Strictly proper scoring rules (SPSR)~\citep{brier1950verification,Win:69,Savage:71,Jose:06, Gneiting:07}  have been developed to elicit private information (e.g. probability assessment about whether the S\&P 500 index will go up next week) and evaluate the reported information for settings where the principal will have access to the ground truth (e.g. whether S\&P 500 index actually went up) at some point.
The score of an agent measures the quality of her prediction. Moreover, facing a strictly proper scoring rule, the agent strictly maximizes her expected score by truthfully revealing her prediction. In this paper, we focus on extending the literature of SPSR to the information elicitation {\em without} verification (IEWV) settings where the principal does not have access to the ground truth and still wants to elicit private probabilistic beliefs. We ask the following question:
\begin{center}
\emph{Can we extend SPSR to scoring mechanisms that can quantify the quality of \\elicited probabilistic information and achieve truthful elicitation for IEWV?} 
\end{center}
We provide a positive answer to this question for multi-task information elicitation. We develop a family of scoring mechanisms that under certain assumptions can estimate a biased version of the ground truth and score predictions against it by removing the bias. As a consequence, we achieve a certain form of dominant truthfulness in eliciting private probabilistic information, a favorable property to have for IEWV \citep{de2016incentives,gao2016incentivizing,kong2016equilibrium,goel2018deep,kong2019information,kong2020dominantly}. To the best of our knowledge, this is the first work to provide a meta solution framework that enables applications of a SPSR to the IEWV setting for eliciting probabilistic beliefs. We name our solution as \emph{Surrogate Scoring Rules}.

As a building block, we first introduce SSR for a stylized setting where the principal has a noisy ground truth (and its error rates) to evaluate the quality of elicited information. We show that SSR preserve the same information quantification and truthful elicitation properties just as SPSR, despite the lack of access to the exact ground truth. 
These surrogate scoring rules are inspired by the use of surrogate loss functions in machine learning~\citep{angluin1988learning,bylander1994learning,scott2013classification,natarajan2013learning,scott2015rate}. They remove bias from the noisy ground truth such that in expectation a report is as if evaluated against the ground truth.

Built upon the above bias correction step, when the principal only has access to agents' reports and one bit of information about the marginal distribution of the ground truth over the entire task set, we develop a multi-agent, multi-task mechanism, \emph{\SSRM}, to again achieve information quantification and truthful elicitation under dominant strategy, when agents adopt the same (arbitrary) strategy for all the tasks they are assigned, and when the principal has sufficiently many tasks and agents. The method relies on an estimation procedure to accurately estimate the average bias in the peer agents' reports. With the estimation, a random peer agent's report serves as a noisy ground truth and SSR can then be applied smoothly to achieve the two desired properties.

We evaluate the empirical performance of SSR with 14 real-world human forecast datasets. The results show that SSR effectively recover, from only agents' reports, the true scores of agents given by SPSR with ground truth. 

We summarize our contributions as follows:
\squishlist
    \item We extend Strictly Proper Scoring Rules (SPSR) to a family of scoring mechanisms, \emph{Surrogate Scoring Rules} (SSR), that operate in the information elicitation without verification (IEWV) setting. SSR only require access to peer reports and one-bit information on the prior, and are able to truthfully elicit probabilistic beliefs.

    \item SSR can build upon any existing  SPSR and quantify the accuracy or value of the reported information as the SPSR do. Therefore, our work complements the proper scoring rule literature, and this extension largely expands the application of SPSR in challenging elicitation setting where the ground truth is unavailable. 
    \item   
    For the IEWV setting, a SSR alike mechanism (\emph{SSR mechanism}) induces dominant truthfulness in reporting. To the best of our knowledge, it is the first dominantly truthful mechanism that elicit probabilistic predictions.\footnote{The mechanism proposed in~\cite{kong2018water} elicits probabilistic predictions but it is not dominantly truthful. The (variants of) mechanisms proposed in~\cite{dasgupta2013crowdsourced,shnayder2016informed,kong2019information,kong2020dominantly} are dominantly truthful but they elicit categorical information.} Therefore, we also contribute to the peer prediction literature via providing a mechanism that elicits truthful probabilistic report in \emph{dominant strategy} and rewards agents according to \emph{prediction accuracy w.r.t.  SPSR} instead of correlation.
    \item We evaluate the empirical performance of SSR mechanism on 14 real-world human prediction dataset. The results show that SSR are able to better assess the true accuracy of agents than other existing peer prediction methods.
\squishend

\paragraph{\bf Organization.} The rest of the paper is organized as follows. We survey the most relevant results in the rest of this section. Section \ref{sec:pre} lays out the preliminaries. 
Section~\ref{sec:model} provides our model of IEWV.
In Section~\ref{sec_IEWNGT}, we study the information elicitation problem in the stylized setting, where there is a noisy version of the ground truth with known bias. We introduce surrogate scoring rules as a powerful solution in this section. In Section~\ref{sec:prob}, we propose the dominantly truthful mechanism, SSR mechanism, to address the general IEWV problem. We present our experimental study about our mechanisms in Section~\ref{sec:exp}. We conclude the paper with Section \ref{sec:con}. 
Missing details and proofs can be found in the Appendix.

\subsection{Related work}

The most relevant literature to our paper is \emph{strictly proper scoring rules} and \emph{peer prediction}. SPSR are designed to elicit subjective beliefs of random variables when the
principal can evaluate agents' prediction after the random variables realize. The pioneer work \citep{brier1950verification} proposes the famous Brier score to quantify the quality of forecasts. Works for variants and full characterization results of SPSR include \citep{Win:69,Savage:71,Jose:06,Gneiting:07}.

Peer prediction is the most popular solution to IEWV. Its core idea is to score each agent based on a reference report elicited from the rest of the agents, and to leverage on the stochastic correlation between different agents' information. 
Earlier peer prediction mechanisms incentivize truthfully reporting at a Bayesian Nash Equilibrium (BNE) \citep{Prelec:2004,MRZ:2005,witkowski2012robust,radanovic2013,Witkowski_hcomp13}.  
Recent works~\citep{dasgupta2013crowdsourced,shnayder2016informed,kong2016equilibrium} have made truthful equilibrium focal in the sense that it leads to the highest expected payoff to agents among all equilibria. But there is at least one other equilibrium that gives the same expected payoff to agents. 
Several more recent works established dominant truthfulness \citep{de2016incentives,gao2016incentivizing,kong2016equilibrium,goel2018deep,kong2019information,kong2020dominantly}. In particular, \cite{kong2016equilibrium,radanovic2016incentives,kong2019information} achieve truthful reporting in dominant strategy with infinite number of tasks, with the follow-up work~\cite{kong2020dominantly} achieving this goal with finite tasks.

Most of the peer prediction works focus on eliciting categorical signals instead of probabilistic beliefs. \cite{kong2018water} provides a mechanism to elicit probabilistic predictions, but truthfully reporting is an equilibrium strategy instead of a dominant strategy.
When the principal does not have the access to the ground truth but an unbiased estimator, \cite{witkowski2017proper} develops a family of proper scoring rules that quantifies the value of probabilistic predictions up to an affine transformation~\cite{frankel2019quantifying}. In comparison, our mechanism does not require to know the ground truth or an unbiased estimate, while it elicits truthful probabilistic predictions in dominant strategy, and qualifies the value of information in the predictions as the SPSR does. We emphasize again that our solution SSR provide a meta framework that maps each existing SPSR to a scoring method to elicit continuous probabilistic predictions.  

As mentioned, our work borrows ideas from the machine learning literature on learning with noisy data (e.g., \cite{natarajan2013learning,frenay2014classification,scott2015rate,van2015learning}). At a high level, our goal in this paper aligns with the goal in learning from noisy labels -- both aim to evaluate a prediction when the ground truth is missing, but instead a noisy signal of the ground truth is available. Our work addresses the additional challenge that the error rate of the noisy signal remains unknown a priori. 


\section{Preliminaries}\label{sec:pre}

Before we introduce our model of information elicitation without verification, we first briefly introduce strictly proper scoring rules (SPSR), which are designed for the well-studied information elicitation with verification settings. We highlight two nice properties of SPSR: (1) SPSR quantify the value of information and (2) SPSR is incentive compatible for elicitation. Our goal of this paper is to develop scoring rules that match these properties for the more challenging without verification settings. Our solutions build upon the understanding of SPSR.

SPSR are designed for eliciting subjective probability distributions of random variables when the principal can reward agents after the realization of the random variables. SPSR apply to eliciting predictions for any random variables, but we introduce them for binary random variables in this section because the rest of our paper focuses on the binary case. 
Let $y\in\{0,1\}$ represent a binary event. An agent has subjective belief $p$ for the likelihood of $y=1$. When the agent reports a prediction $q$ for outcome $y=1$, the principal rewards the agent using a scoring function $S(q, y)$ that depends on both the agent's report and the realized outcome. Strict properness of $S(\cdot, \cdot)$ is defined as follows.
\begin{definition}
A function $S: [0,1] \times \{0,1\} \rightarrow \mathbb R$ that maps the reported belief $q$ and the ground truth $y$ into a score is a \emph{strictly proper scoring rule} if it satisfies \underline{$\mathbb E[S(p, y)] > \mathbb E[S(q, y)]$},
for all $p, q\in[0,1] \text{ and }p\ne q$. The 
expectation is taken with respect to $y\sim\text{Bernoulli}(p)$.
\end{definition}
There is a rich family of strictly proper scoring rules, including Brier ($S(q, y)=1-(q-y)^2$), logarithmic ($S(q, y)=\log(q)$ if $y=1$ and $S(q, y)=\log(1-q)$ if $y=0$) and spherical scoring rules~\citep{Gneiting:07}. 

\paragraph{\bf Incentive compatibility of SPSR} The definition of SPSR immediately gives incentive compatibility. If an agent's belief is $p$, reporting it truthfully uniquely maximizes his expected score.

\paragraph{\bf SPSR quantify value of information} Another nice property of SPSR is that they quantify the value/accuracy of reported predictions. To give a rigorous argument, 
we use an indicator vector $\mathbf{y}$ of length 2 to represent outcome $y$, with 1 at the $y$-th position and 0 otherwise. That is, $\mathbf{y} = (0, 1)$ if $y=1$ and $\mathbf{y} = (1, 0)$ if $y =0$. We use a probability vector $\mathbf{q}=(1-q, q)$ to represent probability $q$. 
By the representation theorem \citep{McCarthy:56,Savage:71, Gneiting:07}, any strictly proper scoring rule can be characterized using a corresponding strictly convex function $G$ as follows:
$
S(q,y) = G(\mathbf{y}) - D_{G}(\mathbf{y},\mathbf{q}),
$
where $D_{G}$ is the Bregman divergence function of $G$.
Now consider the unknown true distribution of $y$, denoted $\mathbf{p^*}=(1-p^*, p^*)$. The expected score (with respect to $\mathbf{p^*}$) of an agent with prediction $q$ is
\[
\mathbb E_{y\sim\mathbf{p^*}}[S(q,y)] = \mathbb E_{y\sim\mathbf{p^*}} [G(\mathbf{y})] - \mathbb E_{y\sim\mathbf{p^*}}[D_{G}(\mathbf{y},\mathbf{q})].
\]
This means that the maximum score an agent can receives in expectation is $\E_{y\sim\mathbf{p^*}} [G(\mathbf{y})]$ and this happens when the agent reports $\mathbf{q} = \mathbf{p^*}$. Moreover, a prediction $\mathbf{q}$ with smaller divergence $\E_{y\sim\mathbf{p^*}}[D_{G}(\mathbf{y},\mathbf{q})]$ receives higher score in expectation. 
Intuitively, $\E_{y\sim\mathbf{p^*}}[D_{G}(\mathbf{y},\mathbf{q})]$ characterizes how ``far away" $\mathbf{q}$ is from the true distribution of $y$ under divergence function $D_G$. This implies that a strictly proper scoring rule $S$ qualifies the the accuracy of a prediction $q$ based on the corresponding divergence function. When $S$ is taken as the Brier scoring rule, the corresponding Bregman divergence is the quadratic function. Then $\E_{y\sim\mathbf{p^*}}[D_{G}(\mathbf{y},\mathbf{q})] = ||\mathbf{p^*}-\mathbf{q}||^2$, implying that a prediction closer to $\mathbf{p^*}$ according to $\ell_2$ norm receives a higher score in expectation. When $S$ is taken as the log scoring rule, the corresponding Bregman divergence is the KL-divergence, $D_{KL}$, which is also called relative entropy. Then, $\E_{y\sim\mathbf{p^*}}[D_{G}(\mathbf{y},\mathbf{q})] = D_{KL}(\mathbf{p^*}||\mathbf{q})+ H(\mathbf{p^*})$ where $H$ is the entropy function. A prediction with smaller KL-divergence from $\mathbf{p^*}$ receives a higher score in expectation. This property of SPSR allows the principal to take an expert's average score over a set of prediction tasks as a proxy of his average accuracy and rank experts accordingly.


\section{Our model}\label{sec:model}

 The goal of this work is to develop scoring mechanisms that quantify the value of elicited information and are incentive compatible, similar to SPSR, but for settings without verification, i.e. when the principal does not have access to the realization of the predicted binary events. We model the information elicitation without verification problem for a multi-task setting. The details of our model and our design goals are described below. 
 
 \subsection{Model of Information Structure}
A principal has a set of $[M]=\{1,...,M\}$ binary random variables (tasks) $y_k \in \{0, 1\}$ for all $k\in [M]$, which she wants to obtain predictions for. Part of our results can be generalized to non-binary tasks, which can be found in Section~\ref{sec_non-binary} of the Appendix. There is a set $[N]=\{1,...,N\}$ of agents. Neither the principal nor the agents have access to the ground truth $y_k$, but agents each observe a private signal $o_{i,k}$, which relates to $y_k$, for task $k$, where $o_{i,k}$ comes from a finite domain $[O_i]=\{0, 1, ..., O_{i}\}$. We allow that the domains of signals differ across agents. We make a few assumptions on the information structure of this setting. 
\begin{assumption}
\label{ass_joint}
Tasks are independent and similar a priori, that is, the joint distribution of $(o_{1, k},...,o_{N, k},y_{k})$ is i.i.d. for all task $k\in[M]$.
\end{assumption}
This assumption is natural when the set of tasks are of similar nature, for example, tasks asking about the reproducibility of studies published in a particular journal within a certain time period. While researchers may a priori hold some beliefs about the journal-wide replication rate, they receive private signals about each study which allows them to give more informed predictions for individual studies. We note that most studies in the field of IEWV make a similar assumption.\footnote{ In~\cite{dasgupta2013crowdsourced,shnayder2016informed,radanovic2016incentives,kong2019information,kong2020dominantly}, where they consider information elicitation for subjective questions (i.e., questions with no ground truth concept, e.g., how do you rank the movie), the authors all assumed that the joint distribution of agents' signals is the same for each task and signals are independent across tasks. In~\cite{kong2018water,kong2020dominantly}, where they consider information elicitation for objective questions (i.e., questions with ground truth), the authors all assumed that the joint distribution of agents' signals together with the ground truth is the same for each task, and all signals and the ground truth are independent across tasks.} 

 Agents share a common prior $p:=\Pr[y_k=1]$ for each task $k$. We denote the distribution of a signal $o_{i,k}$ conditioned on $y_k$ by $\D_{i}^+$ (conditioned on $y_k=1$) and $\D_i^-$ (conditioned on $y_k=0$).  According to Assumption~\ref{ass_joint}, this conditional distribution $(\D_{i}^+,\D_{i}^-)$ is shared across different tasks for agent $i$. We assume that $\D_{i}^+\ne \D_i^-$, otherwise, $o_{i,k}$ is independent with $y_k$. Each agent knows her own $\D_i^+$ and $\D_i^-$.\yc{Am I correct that we only assume that each agent knows her own conditional signal distributions but not those of the other agents?}\jt{correct} For each task, we further assume that agents' signals are independent conditioned on the ground truth. 

\begin{assumption}
\label{ass_cond}
For each task, the agents' signals are mutually independent conditional on the ground truth. That is, $\forall k\in[M], \Pr\left[o_{1,k},...,o_{N,k}|y_k\right]=\prod_{i\in[N]}\Pr[o_{i,k}|y_k]$.
\end{assumption}
This assumption is to exclude scenarios where agents have some form of ``side information'' to coordinate reports. With ``side information'', it is impossible to have any mechanism that can truthfully elicit agents' predictions without access to the ground truth. This issue has been noted in IEWV for objective questions by Kong et al.~\cite{kong2018water,kong2020dominantly} and the same assumption has been adopted. \yc{Why Kong et al. need this assumption for the with verification setting?} \jt{I phrased the sentence in a wrong way, so the meaning is changed. I just modified the sentence. }

Each agent forms her own belief about $y_k$ based on her received signal $o_{i,k}$. We use $p_{i,k}:=\Pr[y=1|o_{i,k}]$ to represent agent $i$'s posterior belief on task $k$. The principal, who knows neither the prior $p$ nor the conditional signal distributions $\D_i^+$ and $\D_i^-$, hopes to elicit predictions $p_{i,k}$ from some agents. \yc{Hmm, the principal doesn't need to get predictions for every single agent, right?}\jt{Yes.} We make a technical assumption about the prior and the knowledge of the principal. 
\begin{assumption}
\label{ass_principal}
The common prior $p\ne 0.5$ and the principal knows $\mathbbm{1}(p>0.5)$.
\end{assumption}
We assume that the principal knows one bit of information about the prior of tasks. This bit of information can help the principal distinguish between a set of truthful predictions vs. a set of inverted predictions (i.e. everyone reporting $1-p_{i,k}$ instead of $p_{i,k}$), which otherwise is impossible. In practice, this bit of information is usually easy to get. For example, the principal may not know the replication rate of a journal but knows whether on average more than half of the studies are successfully replicated. The assumption $p \ne 0.5$ is a technical condition we will need later to distinguish the true scenario from the inverted one.\yc{Add a comment on $p\ne 0.5$?}

$p_{i,k}$  encodes the randomness of $o_{i,k}$. And, $p_{i,k}$ is a discrete random variable with values taken in [0,1].  
Assumptions~\ref{ass_joint} and \ref{ass_cond} jointly imply that the agents' posterior beliefs $p_{i,k}$ are homogeneous across tasks and conditionally independent across agents.  
\begin{proposition}\label{prop_posterior}
Under Assumptions \ref{ass_joint} and \ref{ass_cond}, agents' beliefs $p_{i,k}$ are 
\begin{itemize}
\item Conditionally homogeneous and independent across tasks: For each agent $i\in[N]$, conditioned on $y_k$, her posterior beliefs $p_{i,k}$ are i.i.d. for all tasks $k \in [M]$. That is, $\forall k,k'\in[M]$ and $k\ne k'$, $\forall u\in [0,1], \forall v\in\{0,1\}$, \underline{$\Pr[p_{i,k}=u|y_k=v]=\Pr[p_{i,k'}=u|y_{k'}=v]$}; and $\forall M'\subseteq[M],~\underline{\Pr[\{p_{i,k}\}_{k\in M'}|\{y_k\}_{k\in M'}] = \prod_{k\in M'}\Pr[p_{i,k}|y_k]}$. \yc{Juntao, this condition doesn't look right to me. What's the role of $k$, $k'$ and $v$? They are not really used. Could you please fix the condition?}\jt{Missing $\Pr[p_{i,k}=u|y_k=v]=\Pr[p_{i,k'}=u|y_{k'}=v]$. Added.}
\item Conditionally independent across agents: $\forall k\in[M], \Pr[p_{1,k},...,p_{N,k}|y_k]=\prod_{i\in[N]}\Pr[p_{i,k}|y_k]$. 
\end{itemize}
\end{proposition}
The ``conditionally homogeneous" condition simply states that agent's ``expertise levels" are similar across tasks with same outcomes. In fact, our results hold for models with more general information structures as long as Proposition \ref{prop_posterior} and Assumption \ref{ass_principal} are satisfied.\footnote{Here we allow the priors of different tasks to be different and the $p$ in Assumption~\ref{ass_principal} refers to the mean prior of all tasks.}

\subsection{Mechanism design goals}\label{sec:goals}

The principal is interested in designing a scoring mechanism to facilitate the elicitation of predictions for $y_k$. For each task $k$, the principal can ask some subset $[N_k] \subseteq [N]$ agents to give a prediction $q_{i,k}, \forall i \in [N_k]$. $q_{i,k}$ can be different from $p_{i,k}$. The principal then pays each agent scores based on the predictions she collects from all tasks. We denote $[M_i]\subseteq{M}$ the set of tasks agent $i$ answers. 

Given a mechanism, an agent may report her belief via some strategy and influence the final predictions elicited.
We consider that agents adopt strategies for each task independently, but each strategy could be a mixed strategy. 
\begin{definition}
Let $\Delta_{[0,1]}$ be the space of all probability distributions over $[0,1]$. The strategy of an agent $i$ on task $k$ is a mapping $\sigma:[0,1]\rightarrow \Delta_{[0,1]}$ that maps her posterior belief $p_{i,k}$ into a distribution $\sigma(p_{i,k})$ over [0,1] such that the agent draws a report $q_i$ from $\sigma(p_{i,k})$.
\end{definition}
We define a strategy as a mapping from the space of posterior beliefs, rather than from the space of private signals. This is without loss of generality because if two realizations of $o_{i,k}$ give the same posterior, we can merge the two realizations into one combined realization in our model. 
We also assume that each agent adopts the same strategy across tasks.
\begin{assumption} (Consistent Strategy) 
\label{ass_strategy}
For any agent $i\in[N]$, she adopts the same strategy $\sigma_i(\cdot)$ over all tasks $k\in[M_i]$.
\end{assumption}

This assumption is reasonable as we assume that tasks are a priori similar to each agent. 
We denote the strategy adopted by agent $i$ on all tasks by $\sigma_i(\cdot)$ and denote the  strategy profile of all agents except agent $i$ by $\sigma_{-i}$. We also sometimes abuse our notations and use $\sigma_i$ and $\sigma_{-i}$ to represent the predictions resulted from these strategies.

The principal would like to design a mechanism $\M$ that, when only having access to the reported predictions of the agents, can score agents for each of their reported predictions. The score that agent $i$ receives for predicting $q_{i,k}$ for task $k$, when other agents use strategies $\sigma_{-i}$ on all assigned tasks, is denoted as $R_i(q_{i,k}; \sigma_{-i})$. $R_i(q_{i,k}; \sigma_{-i})$ depends on agent $i$'s prediction on task $k$ and can depend on other agents' predictions on all other tasks. We restrict our attention to anonymous mechanisms and hence drop the subscript $i$ in the score function: we have $R(q_{i,k}; \sigma_{-i})$ as the score of prediction $q_{i,k}$. $\E [R(q_{i,k}; \sigma_{-i})]$ is the expected score that agent $i$ receives for reporting $q_{i,k}$ when other agents use strategies $\sigma_{-i}$. The expectation is taken over the randomness in the ground truth, other agents' signals, and other agents' strategies. 

In this IEWV setting, the principal hopes to design $\M$ with similar properties as what SPSR have for the information elicitation with verification settings: quantification of the value of information and incentive compatibility. 

\paragraph{\bf Quantify value of information}
The score of each prediction should reflect the true accuracy of the prediction, similar to what SPSR achieve. That is, for all $i$, $k$ and $q_{i,k}$ and for any true distribution of ground truth $y_k$, \underline{$\E [R(q_{i,k}; \sigma_{-i})]=f\left(E_{y_k}[S(q_{i,k},y_k)]\right)$} holds for a SPSR $S(\cdot,\cdot)$ and a strictly increasing function $f$.

This design goal aspires that the score an agent receives for a prediction in IEWV recovers what the agent would receive with a SPSR (with access to the ground truth) in expectation.

\paragraph{\bf Dominant truthfulness.} A mechanism is dominantly truthful if each agent reporting truthfully on each assigned task leads to higher expected payoff than other strategies, regardless of other agents' reporting strategies.

\begin{definition}
For an agent $i$, a strategy $\sigma_i$ is a (weakly) dominant strategy if $\forall k \in [M_i]$ and $o_{i,k}$, $\forall i\in[N]$, $\forall \{\D_j^+,\D_j^-\}_{j\in[N]}$, $\forall \sigma'_i, \forall \sigma_{-i}: 
\E [R(\sigma_i; \sigma_{-i})|o_{i,k}]\ge\E [R(\sigma'_i; \sigma_{-i})|o_{i,k}]$,
and $\sigma_i$ is a strictly dominant strategy if the equality holds only when $\sigma'_i=\sigma_i$. 
\end{definition}

A dominant truthful mechanism in IEVW is a mechanism where truthful reporting is each agent's weakly dominant strategy and a strictly dominant strategy if her peers' reports are informative
\footnote{Usually, in a dominant truthful mechanism, truthful reporting is the strict dominant strategy. In IEWV, however, if all the peer agents report predictions independently w.r.t. the ground truth, then there will be no information available for the mechanism to incentivize truthful reporting. Therefore, it is inevitable to allow a dominant truthful mechanism in IEWV to pay truthfully reporting strictly higher only when the peer reports are informative about the ground truth. For example, in~\cite{kong2019information,kong2020dominantly}, the dominant truthful mechanism is defined to be a mechanism that pays truthful reporting strictly higher when for each agent, there exists at least one peer agent reporting truthfully. We will see later that in our definition, we do not require that at least one peer agent reports truthfully. We allow all peer agents to be non-truthful but the mean of their peers reports should be dependent with the ground truth.}. Let $\sigma_i^*$ be the truthful reporting strategy for agent $i$, i.e., $\sigma_i^*$ is the function that maps a belief $p_i$ to a distribution where all probability mass is put on $p_i$. Let \underline{$\bar{q}_{-i,k}:=\frac{1}{N-1}\sum_{j\ne i}q_{j,k}$} be the mean of agents' reported predictions other than agent $i$'s.  Note that $\bar{q}_{-i, k}$ is a random variable because of the randomness in reporting strategy $\sigma_j$ and the randomness in signal $o_{j,k}$ received by agent $j$ for $j\ne i$. We say that $\bar{q}_{-i,k}$ is informative about the ground truth if  $\E[\bar{q}_{-i,k}|y_k=1]\ne\E[\bar{q}_{-i,k}|y_k=0]$. We formally define the dominantly truthful mechanisms as follows.

\begin{definition}
\label{def_dominant}
(Dominant truthfulness). A mechanism $\M$ is \emph{dominantly truthful} if 
$\forall i\in[N],$ $\forall k \in [M_i]$ and $o_{i,k}$, $\forall \{\D_j^+,\D_j^-\}_{j\in[N]},$ $\forall \sigma_i\ne\sigma_i^*, \forall \sigma_{-i}: 
\E [R(\sigma^*_i; \sigma_{-i})|o_{i,k}]\ge\E [R(\sigma_i; \sigma_{-i})|o_{i,k}]$,
and the inequality holds strictly for any strategy profile $\sigma_{-i}$ under which $\bar{q}_{-i,k}$ is informative about $y_k$.
\end{definition}

In Definition~\ref{def_dominant}, we characterize the condition that peers' reports are informative by  that the expectation of the mean of peers' reports differs for different realizations of the ground truth.

\section{Elicitation with noisy ground truth}
\label{sec_IEWNGT}
Before we develop mechanisms with desirable properties for our general model, we first achieve these desirable properties, in this section, under a very stylized setting: \emph{elicitation with noisy ground truth}. In this setting, we introduce surrogate scoring rules as an effective solution. These scoring rules will be the building blocks of our mechanisms for the general model.  

This stylized setting has only one event $y$ and one agent $i$, who observes a signal $o_{i}$ generated from distribution $\D_i(y)$ and forms the posterior $p_{i}=\Pr[y=1|o_i]$. The principal, although cannot observe $y$, has access to a noisy ground truth $z$ that has two \emph{error rates},  $\pose{z}$ and $\nege{z}$, defined as follows:
\underline{$ \pose{z}:=\Pr[z=0|y=1],\, \nege{z}:=\Pr[z=1|y=0].$}
They are the probabilities that $z$ mismatches $y$ under the two realizations of $y$. 
The principal knows the realization $z$ and $\pose{z}, \nege{z}$. The principal cannot expect to do much if $z$ is independent of $y$. Hence, we assume that $z$ and $y$ are stochastically relevant, an assumption commonly adopted in the information elicitation literature~\citep{MRZ:2005}. 
\begin{definition}
Random variable $z$ is stochastically relevant for random variable $y$ if 
the distribution of $y$ conditioned on $z$ is different for different realizations of $z$.
\end{definition}
The following lemma shows that the stochastic relevance requirement directly translates to a constraint on the error rates, that is, $\pose{z}+\nege{z} \neq 1$. 
\begin{lemma}
\label{STOC:RELEVANT}$z$ is stochastically relevant to $y$ if and only if $\pose{z}+\nege{z} \neq 1$. 
\end{lemma}

The goal of the principal in this setting is to design a scoring rule to elicit the posterior $p_{i}$ truthfully using this noisy ground truth $z$ and the knowledge of error rates $\pose{z}, \nege{z}$.
We define the design space of the scoring rule with noisy ground truth as follows.

\begin{definition}
Given a  noisy ground truth $z$ with error rates $(\pose{z}, \nege{z})\in[0,1]^2$, \emph{a scoring rule with noisy ground truth} is a function $R:[0,1]\times\{0,1\}\rightarrow \mathbb{R}$ that maps a prediction $q_i\in[0,1]$ and a realized noisy ground truth $z\in\{0,1\}$ to a score. The function $R$ can depend on error rates  $(\pose{z}, \nege{z})$.
\end{definition}

Adopting the terminology from the scoring rule literature, we refer to strict properness as the property that a scoring rule with noisy ground truth gives a strictly higher expected score to a truthful report than a non-truthful report. 

\begin{definition}
A scoring rule $R(q_i,z)$ with noisy ground truth $z$ is \emph{strictly proper} if it holds for all realizations of $o_i$ and $p_i=\Pr[y=1|o_i]$, that  $ \forall q_i\in[0,1] (q_i\ne p_i), \E_{z|o_i}[R(p_i,z)] > \E_{z|o_i}[R(q_i,z)].$
\end{definition}

\subsection{Surrogate scoring rules (SSR)}\label{sec:ssr}
In this section, we present our solution, the surrogate scoring rules, for this stylized setting. SSR is a family of scoring rules with noisy ground truth and is strictly proper under mild conditions.

\begin{definition}[Surrogate Scoring Rules]

$\Sc: [0,1] \times \{0,1\} \rightarrow \mathbb R_+$ is a surrogate scoring rule if for some strictly proper scoring rule $S: [0,1] \times \{0,1\} \rightarrow \mathbb R_+$ and a strictly increasing function $f:\mathbb R_+ \rightarrow \mathbb R_+$, it holds for  that
$\forall p_i,q_i,\pose{z},\nege{z}\in[0,1]$ and $\pose{z}+\nege{z}\ne 1$, \underline{$\mathbb E_{z}[\Sc(q_i, z)] = f\left( \mathbb E_y[S(q_i, y)]\right)$}, where $y$ is the ground truth drawn from $\text{Bernoulli}(p_i)$ and $z$ is the noisy ground truth generated by $y$ with error rates $\pose{z},\nege{z}.$
\end{definition}

The above definition seeks a surrogate scoring rule $\Sc(\cdot)$ that helps us remove the bias in $z$ and return us a strictly proper score in expectation. The idea is borrowed from the machine learning literature on learning with noisy data \cite{bylander1994learning,natarajan2013learning,scott2015rate,menon2015learning,van2015learning}. SSR can be viewed as a particular class of proxy scoring rules~\citep{witkowski2017proper}. But the approach of \citep{witkowski2017proper} to achieve properness is to plug in an {\em unbiased} proxy ground truth to a strictly proper scoring rule. SSR on the other hand directly work with biased proxy and the scoring function is designed to de-bias the noise. Easily we have the following strict properness result for SSR:
\begin{theorem}
\label{thm_SSR}
Given an agent's fixed prior $p$ and private signal $o_i$, SSR $R(q_i, z)$ with noisy ground truth $z$ is strictly proper for eliciting the posterior $p_i=\Pr[y=1|o_i]$ if $z$ and $o_i$ are independent conditioned on $y$, and $z$ are stochastically relevant to $y$. 
\end{theorem}

We give an implementation of SSR, which we name as  \SSRalpha: 
\begin{small}
\begin{align}
\Sc(q_i, z=1) &= \frac{(1-\nege{z})\cdot S(q_i, 1) - \pose{z} \cdot S(q_i, 0)}{1-\pose{z}-\nege{z}}, \label{surrogate:score1}\\
\Sc(q_i, z=0) &= \frac{(1-\pose{z})\cdot S(q_i, 0) - \nege{z} \cdot S(q_i, 1)}{1-\pose{z}-\nege{z}}, \label{surrogate:score2}
\end{align}
\end{small}
where $S$ can be any strictly proper scoring rule. 
We note that the knowledge of the error rates $\pose{z}, \nege{z}$ is crucial for defining the above SSR. This SSR function is inspired by Natarajan et al.\citep{natarajan2013learning}. It has the following property:
\begin{lemma}[Lemma 1, \citep{natarajan2013learning}]
\label{noise:learn}
For \SSRalpha:
$\forall q_i, \pose{z},\nege{z}\in[0,1]$ and $\pose{z}+\nege{z}\ne 1, \forall y\in\{0,1\}:$ $\E_{z|y}[\Sc(q_i, z)] = S(q_i, y)$.
\end{lemma}
Intuitively speaking, the linear transform in \SSRalpha  will ensure that in expectation, the prediction $q_i$ is scored as if it was scored against $y$ using a SPSR. This can be proved fairly straightforwardly via spelling out the expectation. Interested readers are also referred to \citep{natarajan2013learning}. We would like to note that other surrogate loss functions designed for learning with noisy labels can also be leveraged to design SSR.
\begin{theorem}\label{THM:MAIN}
\SSRalpha is a surrogate scoring rule and $\forall p_i,q_i,\pose{z},\nege{z}\in[0,1] (\pose{z}+\nege{z}\ne 1)$, $\mathbb E_{z}[\Sc(q_i, z)] =  \mathbb E_y[S(q_i, y)]$, where $y$ is the ground truth drawn from $\text{Bernoulli}(p_i)$ and $z$ is the noisy ground truth generated by $y$ with error rate $\pose{z},\nege{z}.$
\end{theorem}
With Theorem \ref{THM:MAIN} we know that \SSRalpha quantifies the quality of information just as the strictly proper scoring rule $S$ does. Further, \SSRalpha has the following variance:
\begin{theorem}\label{THM:VAR}
Let $p_z := \Pr[z=1]$.  \SSRalpha suffers the following variance:
\begin{align}
&\mathbb E_{z}\bigl[ \Sc(q_i, z) - \mathbb E_{z}[\Sc(q_i, z) ] \bigr]^2 = \frac{2p_z \cdot (1-p_z)}{(1-\pose{z}-\nege{z})^2}\cdot \left(S(q_i,1)-S(q_i,0)\right)^2~.
\end{align}

\end{theorem}

\section{Elicitation without verification}\label{sec:prob}

\yc{Deleted parts about mechanism design goals etc. but I haven't edited this section.} \jt{Edited.}

The results in the previous section are built upon the fact that there exists a noisy copy of the ground truth and we know its error rates. In this section, we apply the idea of SSR to  information elicitation without verification. A reasonable way to do so is to take agents' reports as the source for this noisy reference of the ground truth. Yet the principal cannot assume the knowledge of the noise in agents' reports. We find a way to construct a noisy ground truth from agents' report with estimable error rates. We refer this noisy ground truth as the \emph{reference report}. Applying SSR with this reference report, we can finally get a dominantly truthful mechanism that elicits the information and that the payment of the mechanism also quantifies the value of information of agents' reports as what the SPSR do. We call this mechanism \emph{\SSRM}. We present the sketch of our mechanism in Mechanism~\ref{mec_sketch}.

\begin{algorithm}[t]
\caption{\SSRM (Sketch)\label{mec_sketch}}
\begin{algorithmic}[1]
\State For each task $k$, we uniformly randomly pick at least 3 agents, assign task $k$ to them and collect their reported predictions.
\State For each agent $i$ and each task $k$ she answers, we construct a reference report $z_{i,k}$ using peer agents' reports; estimate the error rates $\pose{z_{i,k}}$ and $\nege{z_{i,k}}$ for $z_{i,k}$. 
\State Pay each agent $i$ for $q_{i,k}$ on task $k$ by SSR $R(q_{i,k}, z_{i,k})$ if $\pose{z_{i,k}}+\pose{z_{i,k}}\ne 1$, and pay 0, otherwise.
\end{algorithmic}
\end{algorithm}

The challenge of designing such a mechanism is to construct such a reference report $z_{i,k}$ in Mechanism~\ref{mec_sketch} and successfully estimate its error rates $\pose{z_{i,k}}, \nege{z_{i,k}}$. In the following sections, we show how to construct such a reference report and how to estimate the error rates.

\subsection{Reference report and its property}
Let $s_{j,k}$ be a binary signal independently drawn from Bernoulli$(q_{j,k})$. We term $s_{j,k}$ the \emph{prediction signal} of agent $j$ on task $k$. 
We construct the reference report $z_{i,k}$ for agent $i$ as follows: \emph{We uniformly randomly pick an agent $j$ from the peer agent set $[N]\backslash\{i\}$, collect her prediction $q_{j,k}$, and draw the prediction signal \underline{$s_{j,k}\sim\text{Bernoulli}(q_{j,k})$}. We use this $s_{j,k}$ as the reference report $z_{i,k}$.} 

Clearly,  conditioned on the reports $q_{j,k},j\in[N]$, the distribution of $z_{i,k}$ is Bernoulli$\left(\bar{q}_{-i,k}\right)$ as we uniformly randomly pick a prediction signal.  
Note that in our model, $q_{i,k}\sim\sigma_{i}(p_{i,k}),i\in[N],k\in[M]$. Due to Proposition~\ref{prop_posterior} and Assumption~\ref{ass_strategy},  $\bar{q}_{-i,k}$ is i.i.d. across tasks $k\in[M]$. Thus,
$z_{i,k},k\in[M]$ have the following two properties.

\begin{lemma}
\label{lem_ref_cond}
$\forall i\in[N],k\in[M],\, z_{i,k}$ is independent to agent $i$'s posterior $p_{i,k}$ conditioned on $y_k$.
\end{lemma}

This property ensures that  $z_{i,k}$ can be used as the conditionally independent noisy ground truth by Theorem~\ref{thm_SSR} and thus, SSR with $z_{i,k}$ is strictly proper for eliciting the posterior belief $p_{i,k}$.

\begin{lemma}
\label{lemma_signal}
For any strategy profile agents play, reference reports of an agent $i\in[N]$ are i.i.d. and have the same error rates w.r.t. the ground truth, i.e., $\forall \sigma_1,...,\sigma_N,\forall i\in[N], \exists \pose{i},\nege{i}\in[0,1],\forall k\in[M]:$ $\Pr[z_{i,k}=0|y_k=1]=\pose{i}, \Pr[z_{i,k}=1|y_k=0]=\nege{i}.$
\end{lemma}
This lemma shows that the error rates of the reference reports for agent $i$ are the same across all tasks. This property makes it possible to estimate the error rates using multi-task data. In the following sections, we introduce the estimation of the error rates and complete our mechanism.

\subsection{Asymptotic setting}
\label{sec:bias:learn}
To better deliver our idea for error rates estimation, we start with an asymptotic setting with infinite amounts of tasks and agents, i.e., $M, N \rightarrow \infty$. We will later provide finite sample justification for our mechanism. 

We focus on estimating the error rates of the reference reports for agent $i$.  Based on Lemma~\ref{lemma_signal}, we can use $z$ to denote the reference report for agent $i$ on a generic task, and we only need to estimate the error rates $\pose{z},\nege{z}$ of $z$. Our estimation algorithm relies on establishing three equations. We show that the three equations, with knowing their true parameters (which is true in the asymptotic setting), together will uniquely define $\pose{z},\nege{z}$. Then, in next section, we argue that in the finite sample setting, with imperfect estimate of parameters from agents' reports, the solution from the perturbed set of equations will approximate the true values of $\pose{z},\nege{z}$, with guaranteed accuracy.

To construct the three equations, we make the following preparation. 
Let \underline{$\S_{-i}:=\{s_{j,k}\}_{j\ne i, k\in[M]}$} be a realization of the prediction signals from all agents except $i$ on all tasks.   
For a single task, we draw three random variables $\za$, $\zb$, $\zc$. $\za$ is a prediction signal uniformly randomly picked from all peer agents' prediction signals on that task. Excluding the picked signal $\za$, we then a uniformly randomly pick a  prediction signal and set it as $\zb$. Finally, we uniformly randomly pick a prediction signal as $\zc$, excluding both $\za$ and $\zb$. $\za,\zb,\zc$ are independent conditioned on the ground truth as agents' reports are conditional independent and we have infinite number of agents. 
Meanwhile, $\za$ has the same error rates with the reference report $z$ as they two come from the same random process. With infinite number of agents, $\zb$ and $\zc$ also have the same error rates as $z$. For the same reason to $z$ (Proposition~\ref{prop_posterior} and Assumption~\ref{ass_strategy}), $\za,\zb,\zc$ each is i.i.d. across tasks. Therefore, with infinite tasks, we can know any statistics about $\za,\zb$ and $\zc$ by counting corresponding frequencies on $\S_{-i}$. We can then establish the following three equations.

\noindent\makebox[\linewidth]{\rule{\textwidth}{0.4pt}}

\textbf{1. First-order equation:} The first equation is based on the distribution $z$.  Let \underline{$\posz{-i}:=\Pr[z=1]$}.
$\posz{-i}$ can  be expressed as a function of $\pose{z},\nege{z}$ via spelling out the conditional expectation: 
\begin{align}
\posz{-i} &= p\cdot\Pr[z=1|y=1] + (1-p)\cdot \Pr[z=1|y=0] =p\cdot (1-\pose{z}) + (1-p)\cdot \nege{z}. \label{eqn:1}
\end{align}

\textbf{2. Matching between two prediction signals:} The second equation is derived from a second order statistics, namely the matching probability. We consider the matching-on-1 probability of two uniformly randomly picked prediction signals $\za,\zb$ (on the same task, but from different peer agents). Denote this probability as
\underline{$
\mat{-i}:=\Pr[z_1 =1, z_2 = 1].
$}
This matching probability can be written as a function of $\nege{z},\pose{z}$:
\begin{align}
 \mat{-i} &= p\cdot \Pr\left[z_1=1,z_2=1|y=1\right] + (1-p)\cdot \Pr\left[z_1=1,z_2=1|y=0\right] \nonumber \\
 &= p\cdot \Pr\left[z_1=1|y=1\right]\cdot \Pr\left[z_2=1|y=1\right]+ (1-p)\cdot \Pr\left[z_1=1|y=0\right]\Pr\left[z_2=1|y=0\right] \nonumber \\
&= p\cdot(1-\pose{z})^2 + (1-p)\cdot (\nege{z})^2. \label{eqn:2}
\end{align}

\textbf{3. Matching among three prediction signals:} The third equation is obtained by going one order higher that, we check the matching-on-1 probability over three prediction signals $\za,\zb,\zc$ drawn randomly from three different peer agents on the same task. Denote this probability as \underline{ $\mathree{-i}:= \Pr[z_1 = z_2 =z_3= 1].$} 
Similarly as Eqn. (\ref{eqn:2}),  
we have:
\begin{align}
 \mathree{-i} = p \cdot (1-\pose{z})^3 + (1-p) \cdot (\nege{z})^3. \label{eqn:3}
\end{align}
\noindent\makebox[\linewidth]{\rule{\textwidth}{0.4pt}}

Notice that all three parameters $\posz{-i},\mat{-i},\mathree{-i}$ can be perfectly estimated using $\S_{-i}$ with infinite number of tasks and agents, yet without accessing any of the ground truth. With the knowledge of these three parameters, we prove the following:
\begin{theorem}\label{THM:UNIQUE2}
 $(p, \nege{z},\pose{z})$ are uniquely identified using Eqn.(\ref{eqn:1}, \ref{eqn:2}, \ref{eqn:3}) under Assumption~\ref{ass_principal}, that is, when $p\ne 0.5$ and the principal knows $\mathbbm{1}(p > 0.5)$. 
\end{theorem}
The solution of Eqn.(\ref{eqn:1}, \ref{eqn:2}, \ref{eqn:3}) can be expressed in closed form, which we present in Mechansim~\ref{p_estimate} in the finite sample setting. Now we have completed our mechanism. 
The full mechanism is presented in Mechanism~\ref{mechanism}. We further show that the three equations are both necessary and sufficient to estimate the error rates:
\begin{theorem}\label{THM:SUFFICIENT}
The higher order ($\geq 4$) matching equations do not bring in additional information. 
\end{theorem}

Theorem~\ref{THM:UNIQUE2} shows that without ground truth data, knowing how frequently human agents reach consensus with each other will help us characterize their (average) subjective biases. Further, it implies that \SSRM is asymptotically (in $M,N$) preserving the information quantification as strictly proper scoring rules do and induces a strictly dominant strategy for agent to report truthfully, when $z$ is informative (weakly dominant strategy otherwise). To see this, because both $\pose{z},\nege{z}$ are set to their true values, we have $\mathbb E [R(q_{i,k},z)] =  \mathbb E[S(q_{i,k},y)]$. Formally,

\begin{algorithm}[t]
\caption{\SSRM}\label{mechanism}
\begin{algorithmic}[1]

\State For each task $k$, uniformly randomly pick at least 3 agents, assign task $k$ to them, collect their reported predictions and generate the prediction signal for each prediction.

\State For each agent $i$ and each task $k$ she answers, uniformly randomly select one prediction signal $s_{j,k}$ from her peers' prediction signals on the same task and let the reference report $z_{i,k}:=s_{j,k}$.

\State Solve Eqn.(\ref{eqn:1}, \ref{eqn:2}, \ref{eqn:3}) to obtain $\nege{z},\pose{z}$.
\State Pay each agent $i$ for $q_{i,k}$ on task $k$ by \SSRalpha if $\pose{z_{i}}+\pose{z_{i}}\ne 1$, and pay 0, otherwise.
\end{algorithmic}
\end{algorithm}

\begin{theorem}\label{THM:ASYM}
When $z$ is informative, asymptotically ($M,N\rightarrow \infty$) the expected score of \SSRM 
equals to the score of its corresponding strictly proper scoring rule $S$:
\underline{$
\mathbb E [R(q_{i,k},z)] = \mathbb E[S(q_{i,k},y)].
$}
\end{theorem}

\begin{corollary}
\label{cor_infinite}
\SSRM is dominantly truthful with infinite number of tasks and agents. 
\end{corollary}

\begin{remark}
Theorem~\ref{THM:UNIQUE2} and~\ref{THM:ASYM} rely on Proposition~\ref{prop_posterior} and Assumptions~\ref{ass_principal} and \ref{ass_strategy}. Proposition~\ref{prop_posterior} and Assumption~\ref{ass_strategy} guarantee that there exists, across the predictions of different tasks, a similar information pattern that we can learn to infer the ground truth. Therefore, they can be hardly relaxed in IEVW settings. 
For Assumption~\ref{ass_principal}, we'd like to argue that at least one bit of information is needed in order to distinguish the case when agents are truthfully reporting from the case that agents are misreporting by reverting their observations. This is because for every possible tuple $(p, \nege{z},\pose{z})$ resulted by truthful reporting from agents, consider the following counterfactual world: relabeling $0 \rightarrow 1$ and $1 \rightarrow 0$, we will have another distribution of observations characterized by the tuple $(1-p, \pose{z},\nege{z})$. Then agents misreporting will lead to a distribution with parameters being the same as $(p, \nege{z},\pose{z})$. Thus the mechanism designer cannot tell the above two cases apart. Some work~\cite{kong2020dominantly} relaxes Assumption~\ref{ass_principal} by excluding the ``relabeling equilibrium'' from consideration.
\end{remark}

We will show in the next section, \SSRM is also dominantly truthful with finite number of tasks and agents under mild conditions. Several remarks follow. (1) We would like to emphasize again that for an agent $i$, both $z$ and $\Sc(\cdot)$ come from prediction signals of her peer agents' reports $\S_{-i}$: $z$ will be decided by agents $j \neq i$'s reports $\S_{-i}$. $\Sc(\cdot)$ not only has $z$ as input, but its definition also depends on $\pose{z}$ and $\nege{z}$, which will be learned from $\S_{-i}$. (2) When making decisions on reporting, we show under our mechanisms agents can choose to be oblivious of how much error presents in others' reports. This removes the practical concern of implementing a particular Nash Equilibrium. (3) Another salient feature of our mechanism is that we have migrated the cognitive load for having prior knowledge from agents to the mechanism designer. Yet we do not assume the designer has direct knowledge neither; instead we will leverage the power of estimation from reported data to achieve our goal.

\subsection{Finite sample analysis}

With finite $M,N$, there are multiple reasons that we won't be able to obtain perfect estimates of $\pose{z},\nege{z}$. 
For instance, in forming Eqn.(\ref{eqn:1}, \ref{eqn:2}, \ref{eqn:3}), the error rates of two randomly picked prediction signals $\zb,\zc$ will not have the exactly same error rates with $z$. 
However when the number of agent is large enough, we will show that the error rates of $\zb,\zc$
can approximate these $\pose{z}, \nege{z}$ with small and diminishing errors (as a function of number of agents $N$). This can factor into the errors in estimating $\mat{-i}$. Furthermore, the algorithm's estimates of the following three parameters for each agent $i$, $\posz{-i},\mat{-i},\mathree{-i}$, are not perfect. 

All three parameters $\posz{-i},\mat{-i},\mathree{-i}$ can be estimated from agents' reports, without the need of knowing any ground truth labels. Let $k_1,k_2,k_3$ be the three agents whose prediction signals are selected as $\za,\zb,\zc$ for each task $k\in[M]$ (In practice, we only need to assign task $k$ to these three randomly selected agents).  
Then we estimate:
\begin{align*}
\widetilde{\posz{-i}} &= \frac{\sum_{k=1}^M \mathbbm{1}(s_{k_1,k} = 1 )}{M},~\widetilde{\mat{-i}} = \frac{\sum_{k=1}^M \mathbbm{1}(s_{k_1,k} = s_{k_2,k} = 1 )}{M},~\widetilde{\mathree{-i}} = \frac{\sum_{k=1}^M \mathbbm{1}(s_{k_1,k} = s_{k_2,k}=s_{k_3,k}= 1 )}{M}.
\end{align*}
We then solve the system of equations (\ref{eqn:1}, \ref{eqn:2}, \ref{eqn:3}) with these estimates to obtain estimated error rates $\pose{z},\nege{z}$. We present the solution in Mechanism~\ref{p_estimate}.
\begin{algorithm}[t]
\caption{ Estimation of $\pose{z},\nege{z}$}\label{p_estimate}
\begin{algorithmic}[1]
\State Estimate  $\widetilde{\posz{-i}}, \widetilde{\mat{-i}}, \widetilde{\mathree{-i}}$. Compute the following quantities:
\[
a = \frac{\widetilde{\mathree{-i}}-\widetilde{\posz{-i}}\widetilde{\mat{-i}}}{\widetilde{\mat{-i}} - (\widetilde{\posz{-i}})^2}, ~b = \frac{\widetilde{\posz{-i}} \widetilde{\mathree{-i}} - (\widetilde{\mat{-i}})^2}{\widetilde{\mat{-i}} - (\widetilde{\posz{-i}})^2},~
\underline{x} =  \frac{a -\sqrt{a^2-4b}}{2}, ~\overline{x} = \frac{a+\sqrt{a^2-4b}}{2}
\]
\State Denote by $\underline{e},\overline{e}$ as the $\underline{x},\overline{x}$ that are closer and further to $\widetilde{\posz{-i}}$ respectively:
\[
\underline{e} = \text{argmin}_{x \in \{\underline{x},\overline{x}\}} |x-\widetilde{\posz{-i}}|,~~~\overline{e} = \text{argmax}_{x \in \{\underline{x},\overline{x}\}} |x-\widetilde{\posz{-i}}|
\]
\State If $p < 0.5$: $\widetilde{\nege{z}}:= \underline{e},~~\widetilde{\pose{z}}:= 1 - \overline{e}$; else if $p > 0.5$: $\widetilde{\nege{z}}:= \overline{e},~~\widetilde{\pose{z}}:=  1 - \underline{e}~.$
\end{algorithmic}
\end{algorithm}

We give a statistical consistency analysis for this estimation procedure for this finite sample setting. We bound the estimation error in estimating reports' error rate as a function of $M$ and $N$. The first source of errors is due to the imperfect estimations of $\mat{-i},\mathree{-i},\posz{-i}$. The second one is due to estimation errors for matching probability with heterogeneous agents. Formally we have the following theorem:

\begin{lemma}
\label{lem_err}
$\widetilde{\pose{z}}, \widetilde{\nege{z}}$ given by Mechanism~\ref{p_estimate} satisfy
$|\widetilde{\pose{z}}-\pose{z}| \leq \epsilon,~|\widetilde{\nege{z}}-\widetilde{\nege{z}}| \leq \epsilon$ with probability at least $1-\delta$, where \underline{$
\epsilon:= O\bigl(\frac{1}{N}+\sqrt{\frac{\ln \frac{1}{\delta}}{M}}\bigr)$}, which can be made arbitrarily small with increasing $M$ and $N$.
\end{lemma}

Denote by $\Delta:=(1-p) (1-\nege{z}-\pose{z})$. The above estimation of $\pose{z},\nege{z}$ further leads to the following the above consistency result:
\begin{theorem}\label{THM:EST}
For the scoring function $\widetilde{R}(\cdot)$ defined for \SSRalpha using $\widetilde{\pose{z}},\widetilde{\nege{z}}$, when $M,N$ are large enough s.t. $\epsilon \leq (1-\nege{z}-\pose{z})/4$, with probability at least $1-\delta$, $$|\widetilde{R}(q_i,z)-R(q_i,z)| \leq \frac{12\epsilon \cdot \max S}{\Delta^2},~\forall q_i\in[0,1],z\in\{0,1\},$$
where $\max S$ is the maximum score of the underlying SPSR that $\widetilde{R}$ builds on. This further implies that 
\begin{align*}
    |\E[\widetilde{R}(q_i,z)]-\E[R(q_i,z)]| \leq \frac{12\epsilon \cdot \max S}{\Delta^2},~ |\E[\widetilde{R}(q_i,z)]-\E[S(q_i,y)]| \leq \frac{12\epsilon \cdot \max S}{\Delta^2},~\forall q_i\in[0,1]
\end{align*}
\end{theorem}

Now we present the incentive guarantees in finite sample regime under noisy estimations. We first note that any linear transformation of a particular \SSRM preserves its incentive property. To simply our analysis, we will first perform the following operation to ``cancel" the effects of noisy estimation of $\pose{z},\nege{z}$ in the denominator of $\Sc(\cdot)$:
\underline{$
\widetilde{R}(q_i,z):= (1-\widetilde{\pose{z}}-\widetilde{\nege{z}})\cdot \widetilde{R}(q_i,z)
$} - note the above linear transform (independent of agent's reports) does not change the incentive property of SSR.
\begin{theorem}\label{FINITE}
When $z$ is informative, set 
$M,N$ large enough but finite, \SSRM returns a score that is $\epsilon(M,N)$ close to the score of its corresponding strictly proper scoring rules, where $\epsilon(M,N)=O\bigl(\frac{1}{N}+\sqrt{\frac{\ln M}{M}}\bigr)$ is a diminishing term in both $M$ and $N$. Further, for each agent $i$,  it is a strictly dominant strategy to truthfully report $q_{i,k}, \forall k$ when $S(q,y)$ is strongly concave and Lipschitz in $q$ for any $y\in\{0,1\}$ and $M,N$ are sufficiently large. 
\end{theorem}
The intuition about dominant truthfulness part is that when $M,N$ are sufficiently large, the estimation error is too small such that the deviation gain through utilizing the error cannot surpass the loss in the true score, and the qualified $M,N$ are determined by the curvature of $S(\cdot)$.

\begin{corollary}
When SPSR $S(q, y)$ is strongly concave and Lipschitz in $q$ for all $y\in\{0,1\}$, the \SSRM built upon $S(\cdot)$ is dominantly truthful with finite but sufficiently large $N$ and $M$. 
\end{corollary}

For example, Log scoring rule over interval $[0.01, 0.99]$ is strongly concave and Lipschitz.\footnote{When log scoring rule is applied, the range of the prediction is usually restricted to a closed interval excluding point 0 and 1, e.g., [0.01, 0.99]. This is because log scoring rule is not well-defined (infinite) when the prediction is 0 (or 1) while the ground truth is 1 (or 0).}

\section{Empirical studies}
\label{sec:exp}
Using 14 real-world human forecasting datasets, we demonstrate that without the need of accessing ground truth, \SSRM demonstrate  stronger correlation with the true scores given by SPSR (which use ground truth outcome) than the other peer prediction methods across different datasets we tested over.

\subsection{Setting}
We evaluate the properties of \SSRM (built upon three popular SPSR) with 14 real-world forecasting datasets and compare the results to those of other four popular existing peer prediction methods. In what follows, we introduce the details of these settings.

\subsubsection{Datasets}
We conduct our experiments on 14 datasets from three human forecasting and crowdscourcing projects: the Good judgment Project (GJP), the Hybrid Forecasting Project (HFC) and an MIT collected human judgment datasets. These three projects are different in both the populations of participants, forecast topics and elicitation methods. 

\paragraph{\textbf{\emph{GJP datasets~\cite{atanasov2016distilling}}}} It contains four datasets on geopolitical forecasting questions. The four datasets, denoted by G1$\sim$G4, was collected from 2011 to 2014 respectively. They have different forecasting questions and forecasters. Each forecaster has a single probabilistic prediction for a question she answered in the datasets.

\paragraph{\textbf{\emph{HFC datasets~\cite{HFC}}}} It contains three datasets, denoted by H1$\sim$H3, collected from the Hybrid Forecast Competition organized by IARPA in 2018. The three datasets share the same forecasting questions about geopolitics, finance, economics, etc, but have different forecasters and collecting methods. 
These three datasets record multiple probabilistic predictions each forecaster made at different dates. We used the final prediction made by a forecaster on a question she answered.

\paragraph{\textbf{\emph{MIT datasets~\cite{prelec2017solution}}}} It contains seven datasets, denoted as M1a, M1b, M1c, M2, M3, M4a, M4b, with different questions and forecasters. The questions ranges from the capital of states to the price interval that artworks belong to, to some trivia questions. The forecasters were students in class and colleagues in labs. In datasets M1a, M1b, M4a, M4b, forecasters made binary vote on a forecasting question. In datasets M1c, M2, M3, forecasters gave a probabilistic prediction. 

\vspace{0.5em}
We focus on the forecasting questions with binary outcomes in these datasets. We filtered out the questions with less than 10 submitted predictions and the participants who predicted on less than 15 questions. No questions were filtered out from GJP and MIT datasets and only a few from HFC datasets.
Basic statistics of these datasets are presented in Table~\ref{tab_datasets_full}. %

\begin{table*}[t]
\scriptsize
\centering
\setlength{\tabcolsep}{4pt}
\begin{tabular}{cccccccccccccccccccccc}
\toprule
\titlecell{c}{Items} & \titlecell{c}{G1} & \titlecell{c}{G2} & \titlecell{c}{G3} & \titlecell{c}{G4} & \titlecell{c}{H1} & \titlecell{c}{H2} & \titlecell{c}{H3} & \titlecell{c}{M1a} & \titlecell{c}{M1b} &\titlecell{c}{ M1c} & \titlecell{c}{M2} & \titlecell{c}{M3} & \titlecell{c}{M4a} & \titlecell{c}{M4b}\\ 
\toprule

\# of questions (original) & 94 & 111 & 122 & 94 & 88 & 88 &  88 & 50 & 50 & 50 & 80 & 80 & 90 & 90\\ 

\# of agents (orginal)  & 1972 & 1238 & 1565 & 7019 & 768 & 678 & 497 & 51 & 32 & 33 & 39 & 25 & 20 & 20\\ 
\cmidrule(lr){1-15}

\multicolumn{15}{c}{After applying the filter}\\

\cmidrule(lr){1-15}
\# of questions & 94  & 111  & 122  & 94  & 72  & 80 & 86 & 50 & 50 & 50 & 80 & 80 & 90 & 90\\

\# of agents & 1409 & 948 & 1033 & 3086 & 484 & 551 & 87 & 51 & 32 & 33 & 39 & 25 & 20 & 20\\


Avg. \# of answers per question & 851 & 533 & 369 & 1301 & 188 & 252 & 33 & 51 & 32 & 33 & 39 & 18 & 20 & 20 \\

Avg. \# of answers per agent & 57 & 62 & 44 & 40 & 28 & 37 & 33 & 50 & 50 & 50 & 80 & 60 & 90 & 90 \\


Majority vote correct ratio (\%)  & 0.90 & 0.92 & 0.95 & 0.96 & 0.88 & 0.86 & 0.92 & 0.58 & 0.76 & 0.74 & 0.61 & 0.68 & 0.62 & 0.72 \\
\bottomrule
\end{tabular}
    \caption{Statistics about binary-outcome datasets from GJP, HFC and MIT datasets}

    \label{tab_datasets_full}
\end{table*}

\subsubsection{SPSR}
We consider three SPSR: Brier score, log scoring rule, and rank-sum scoring rule. 
The first two are the most widely adopted scoring rules, and they are equivalent to squared error and cross-entropy loss, respectively, for measuring the accuracy of predictions. The rank-sum scoring rule can be written as an affine transformation (depending on the number of tasks in each ground truth category) of AUC-ROC metric,   ~\cite{parry2016linear}. Therefore, it is also of interest to us.

In the experiments, we adopt the convention used the in the GJP for Brier score that it ranges from 0 to 2 and a smaller score corresponds to a higher accuracy.\footnote{This is different from using SPSR as a payment method, where the higher the better. We can transfer between these two usages by applying a negative scalar.} To align with Brier score, we also use a log scoring rule and a rank-sum score rule that a smaller score corresponds to a higher accuracy and the minimum possible score is 0. 

Let $[M_i]$ be the set of tasks answered by agent $i$. Recall that $q_{i,k}$ and $y_k$ are agent $i$'s prediction and the ground truth for task $k$, respectively. The exact formulas for the three scoring rules we used are as follows:
\squishlist
    \item \textbf{Brier score}: $S^{\textsf{Brier}}(q_{i,k}, y_k)=(q_{i,k}-y_k)^2+\left((1-q_{i,k})-(1-y_k)\right)^2 = 2(q_{i,k}-y_k)^2.$
    
    An agent's accuracy score under Brier score is the mean Brier score $\frac{1}{M_i}\sum_{k\in [M_i]}S^{\textsf{Brier}}(q_{i,k}, y_k)$.
    
    \item \textbf{Log scoring rule}: $S^{\textsf{log}}(q_{i,k}, y_k) = \log(q_{i,k})$ if $y_k=1$; and $S^{\textsf{log}}(q_{i,k}, y_k) = \log(1-q_{i,k})$ if $y_k=0$.
    
    An agent's accuracy under log scoring rule is also the mean score $\frac{1}{M_i}\sum_{k\in [M_i]}S^{\textsf{log}}(q_{i,k}, y_k)$. As it is unbounded in the worst case, we change all predictions with value 1 to 0.99 and predictions with value 0 to 0.01 to ensure a well-defined score. 
    
    \item \textbf{Rank-sum scoring rule} is a multi-task scoring rule. For a single task $k$, it assigns a score $$S^{\textsf{rank}}(q_{i,k}, y_k)=-y_k \cdot \psi\left(q_{i,k}|\{q_{i,k'}\}_{k'\in[M_i]}\right),$$ where $\psi\left(q_{i,k}|\{q_{i,k'}\}_{k'\in[M_i]}\right):=\sum_{k'\in[M_i]}\mathbbm{1}(q_{i,k'}<q_{i,k})-\sum_{k'\in[M_i]}\mathbbm{1}(q_{i,k'}>q_{i,k})$ is the rank of prediction $q_{i,k}$ in all agent $i$'s predictions. Then, agent $i$'s rank-sum score $S_i^{\textsf{rank}}$ is defined: $S_i^{\textsf{rank}}=\sum_{k\in[M_i]}S^{\textsf{rank}}(q_{i,k}, y_k)$.\footnote{The AUC-ROC of agent $i$ is equal to  $\frac{1}{2}\left(1-\frac{1}{M^+_i(M_i-M^+_i)}S_i^{\textsf{rank}}\right)$, where $M_i+:=\sum_{k'\in[M_i]}\mathbbm{1}(y_{k'}=1)$~\cite{parry2016linear}.}
    The range of the score increases with the number of answered tasks quadratically. We normalize the score using $1+\frac{4}{M_i^2}S_i^{\textsf{rank}}$ to range [0, 2].  
\squishend

\subsubsection{Treatments}
Though existing peer prediction methods are not designed for recovery of SPSR, we add comparisons to them for completeness of our study.\footnote{We do not intend to claim our mechanism is better in any sense, as it would be an unfair comparison since the goals were different in each design of these mechanisms.}
In particular, we'd like to understand whether in practice SSR has the advantage of revealing the true scores given by SPSR while not accessing ground truth information.

In our experiments, we consider four popular existing peer prediction methods, serving as comparisons to SSR: proxy scoring rule (PSR) with extremized mean~\cite{witkowski2017proper}, peer truth serum (PTS)~\cite{radanovic2016incentives}, correlated agreement (CA)~\cite{shnayder2016informed}, determinant mutual information (DMI)~\cite{kong2020dominantly}. 

PSR is to directly apply the SPSR w.r.t. an unbiased proxy of the ground truth, and Witkowski et al recommended using the extremized mean of the reported predictions as the unbiased proxy, when there is no verification data available~\cite{witkowski2017proper}. Using different SPSR as the building block, we can get different PSR. 
PTS, CA, DMI are not build upon SPSR and are designed to elicit a categorical label instead of a probabilistic prediction. When applied them on datasets with probabilistic predictions, we assume that a categorical label is drawn from the probabilistic prediction and we compute an asymptotically consistent estimator of their expected scores, where the expectation is taken over the drawn of the categorical label.  

\subsection{Main results}

\noindent \paragraph{\textbf{\emph{Unbiasedness of SSR}}} We exam to what extend SSR recover the true accuracy scores given by different SPSR. We compute the true mean score and mean SSR score of each human forecaster in all datasets. 

\begin{figure*}[t]
   \centering
    \begin{subfigure}[t]{0.32\textwidth}
        \centering
        \includegraphics[width=\textwidth]{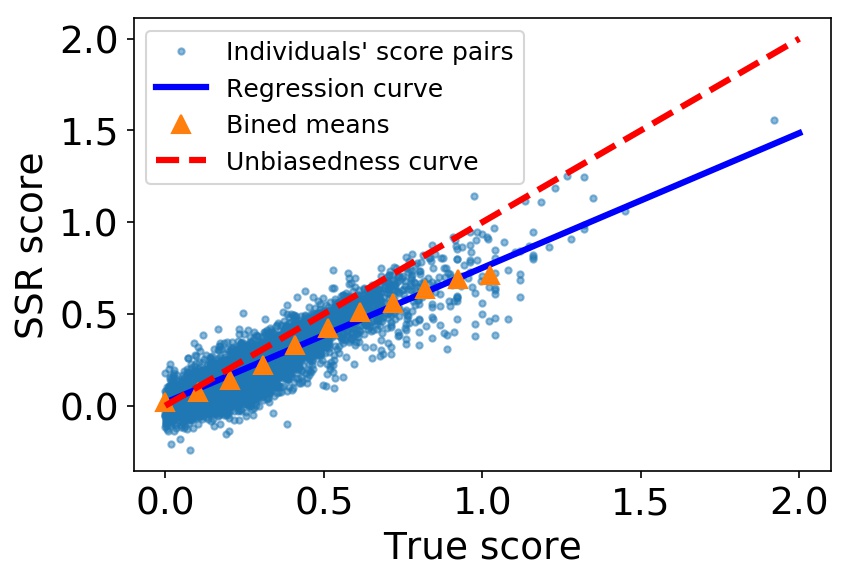}
        \caption{Brier  ($y=0.787 \cdot x + 0.001$)}
    \end{subfigure}  
    \begin{subfigure}[t]{0.32\textwidth}
        \centering
        \includegraphics[width=\textwidth]{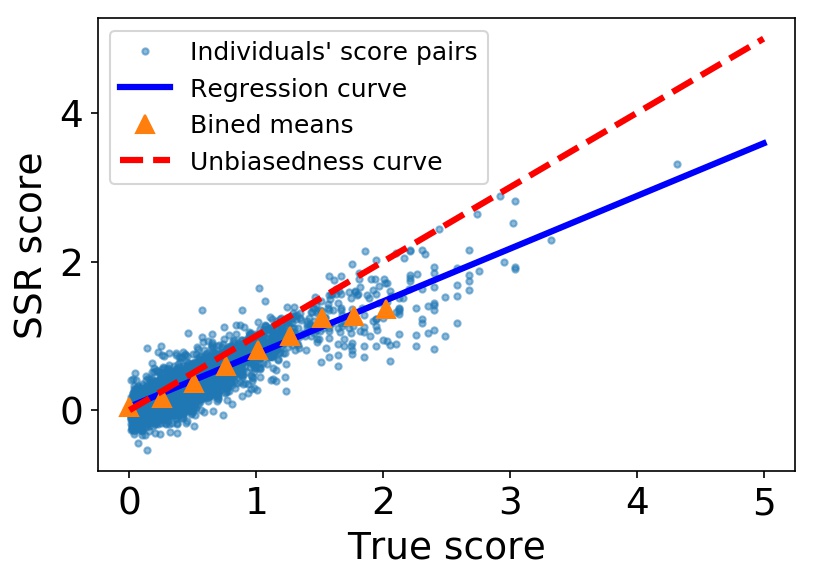}
        \caption{Log ($y=0.790 \cdot x - 0.005$)}
    \end{subfigure}  
    \begin{subfigure}[t]{0.32\textwidth}
        \centering
        \includegraphics[width=\textwidth]{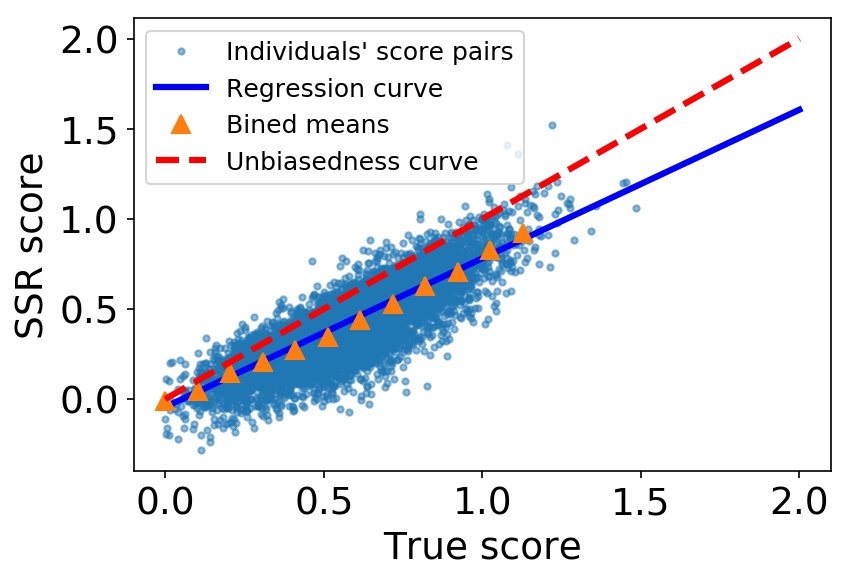}
        \caption{Rank-sum ($y=0.839 \cdot x - 0.057$)}
    \end{subfigure}
    \vspace{-0.7em}
    \caption{Regression of individuals' true accuracy and SSR score over 14 datasets under three different SPSR.\label{fig_unbiase}}
    \vspace{-0.7em}
\end{figure*}

The pairs of true mean accuracy score and mean SSR score of every individual in the 14 datasets are illustrated by blue dots in Fig~\ref{fig_unbiase}. It is clear that most of them concentrate around $y=x$, which demonstrates the unbiasedness of SSR scores. 
Then, we separate forecasters into different bins w.r.t. their true scores. For Brier score and rank-sum scoring rule, the centers of the bins are from 0 to 2.0 with a width of 0.05. For log scoring rule, the centers of the bins are from 0 to 5 with a width of 0.1. 
For forecasters in each bin, we then calculate the mean SSR score of these forecasters (we ignore bins with less than 20 forecasters). We find that for users at same true score level, their SSR scores are also at at similar level. These are illustrated by orange triangles in Fig~\ref{fig_unbiase}.
Finally, we draw the linear regression curves on these binned means such that each true accuracy level is weighted uniformly in the regression (blue curve in Fig~\ref{fig_unbiase}). The slope for the three curves are all around 0.8, while the intercepts are all round 0. This shows that the average SSR score is extremely close to the true accuracy score when the true accuracy score is small. In other words, SSR can calibrate the true accuracy almost perfectly for sophisticated forecasters. 
Given most agents have a true accuracy score better than uniformly randomly guessing 0 and 1 (which is 1 in Brier score and rank-sum score and 2.3 in log score) in these 14 datasets, SSR approximate the true scores well for most of the time.

\begin{figure*}[t]
   \centering
    \begin{subfigure}[t]{0.27\textwidth}
        \centering
        \includegraphics[width=\textwidth]{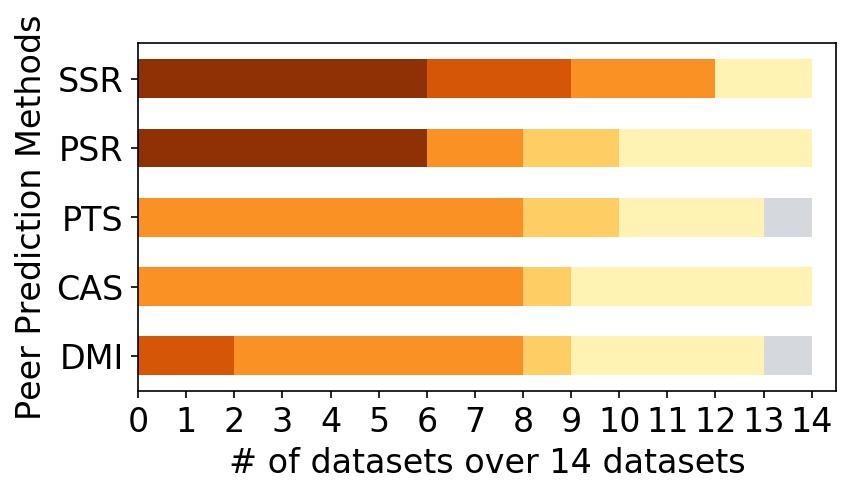}
        \caption{Brier score}
    \end{subfigure}  
    \begin{subfigure}[t]{0.27\textwidth}
        \centering
        \includegraphics[width=\textwidth]{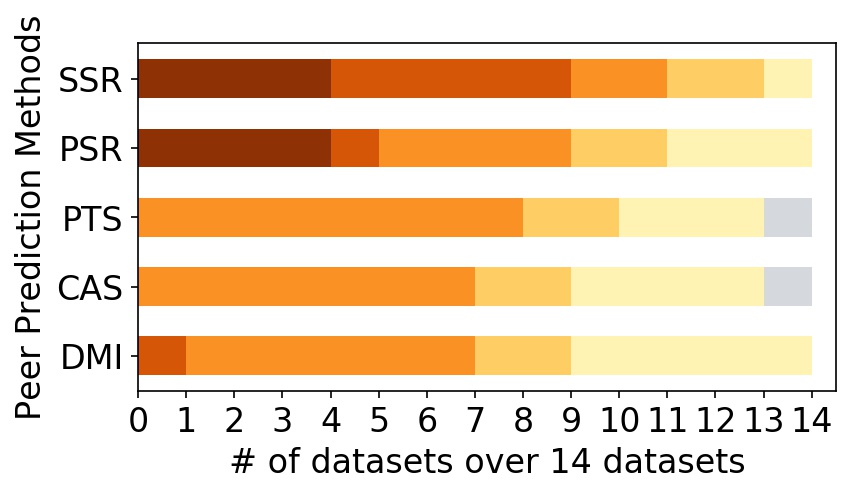}
        \caption{Log scoring rule}
    \end{subfigure}  
    \begin{subfigure}[t]{0.27\textwidth}
        \centering
        \includegraphics[width=\textwidth]{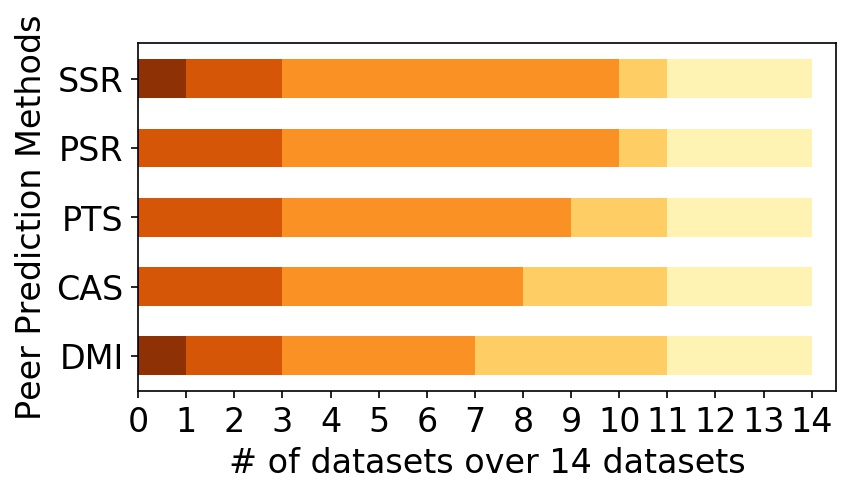}
        \caption{Rank-sum scoring rule}
    \end{subfigure}
    \begin{subfigure}[t]{0.15\textwidth}
        \centering
        \includegraphics[width=\textwidth]{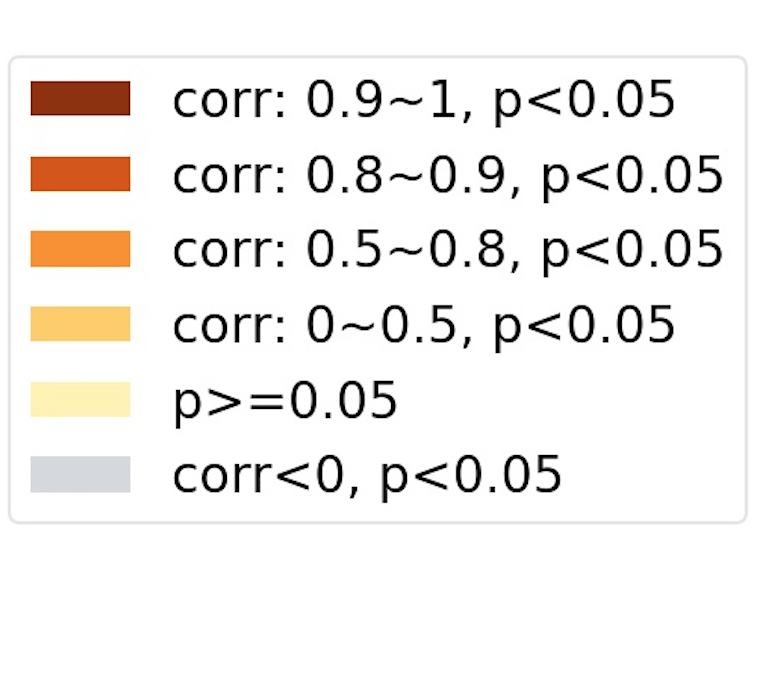}
    \end{subfigure}
    \vspace{-0.7em}
    \caption{The number of datasets in each level of correlation (measured by Pearson's correlation coefficient) between individuals' peer prediction scores and different SPRS.\label{fig_corr_pearson}}
    \vspace{-1em}
\end{figure*}

\paragraph{\textbf{\emph{Correlation with SPSR}}}
We exam the correlations between agents' peer prediction scores and true accuracy scores given by the three SPSR, Brier score, log scoring rule and rank-sum scoring rule. When a SPSR is chosen as the true score, we also use this SPSR as the underlying scoring rule called by SSR and PSR. PTS, CA and DMI scores are independent from which SPSR is used. We adjust the scores such that a lower score corresponds to a higher accuracy (or a higher payment to the agents) in the context of each peer prediction method.

We exam these correlations on each dataset independently, and categorize the level of correlations according to the Pearson's correlation coefficient and p-values. 
As shown in Fig~\ref{fig_corr_pearson}, we find that for Brier score, and log scoring rule, SSR achieves a Pearson's correlation $\textsc{coefficient}>0.8$ on 9 out of 14 datasets. The second best, PSR, achieves a $\textsc{coefficient}>0.8$ on at most 6 out of 14 datasets. PTS and CAS do not have a $\textsc{coefficient}>0.8$ on any datasets, while DMI achieves $\textsc{coefficient}>0.8$ on at most 2 of the datasets. For rank-sum scoring rule, all peer prediction scores achieve similar levels of correlation among 14 datasets, while SSR are better than the others. We observe similar results on Spearman's correlation test (Fig~5 in the Appendix). This result on Spearman's (rank) test, in particular, implies that \SSRM rank the agents in a similar order of agents' true expertise.

\begin{figure*}[t]
   \centering
    \begin{subfigure}[t]{0.3\textwidth}
        \centering
        \includegraphics[width=\textwidth]{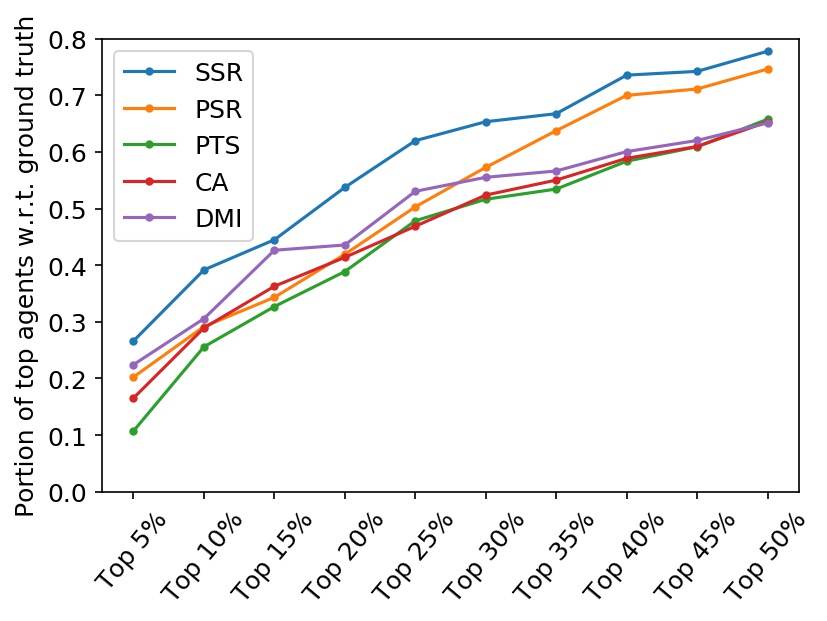}
        \caption{Mean squared loss}
    \end{subfigure}\hspace{1em}
    \begin{subfigure}[t]{0.3\textwidth}
        \centering
        \includegraphics[width=\textwidth]{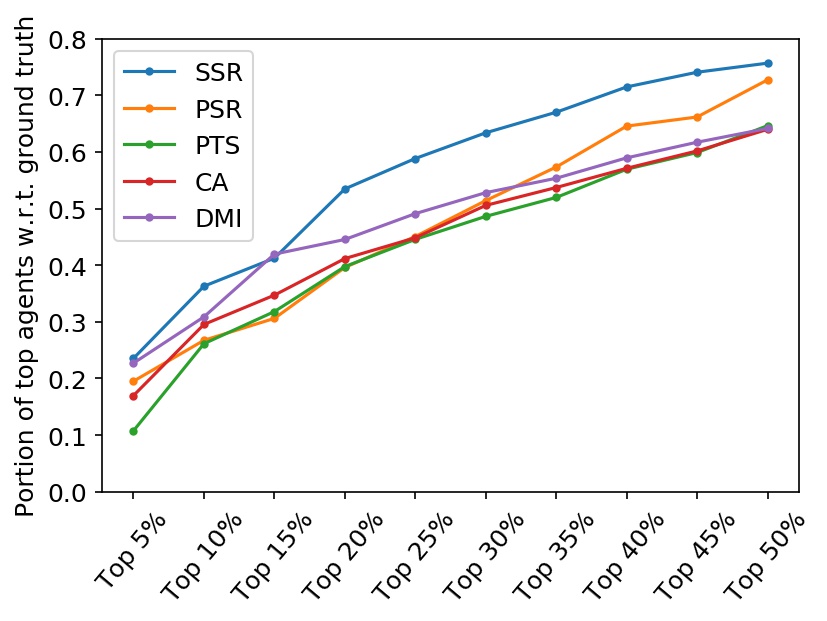}
        \caption{Mean cross-entropy loss}
    \end{subfigure}\hspace{1em}
    \begin{subfigure}[t]{0.3\textwidth}
        \centering
        \includegraphics[width=\textwidth]{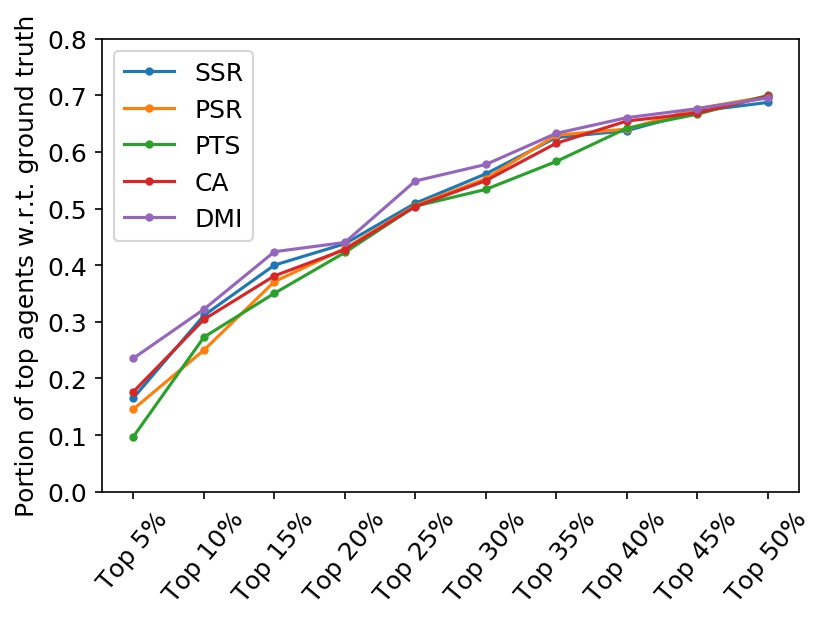}
        \caption{AUC-ROC}
    \end{subfigure}
    \vspace{-0.7em}
    \caption{The portion of top $t$\% forecasters w.r.t. 3 different metrics (mean squared loss, cross-entropy loss, AUC-ROC loss) in the top $t$\% forecasters selected by different methods (averaged over 14 datasets).\label{fig_top_user}}
    \vspace{-1em}
\end{figure*}

\begin{figure*}[t]
   \centering
    \begin{subfigure}[t]{0.3\textwidth}
        \centering
        \includegraphics[width=\textwidth]{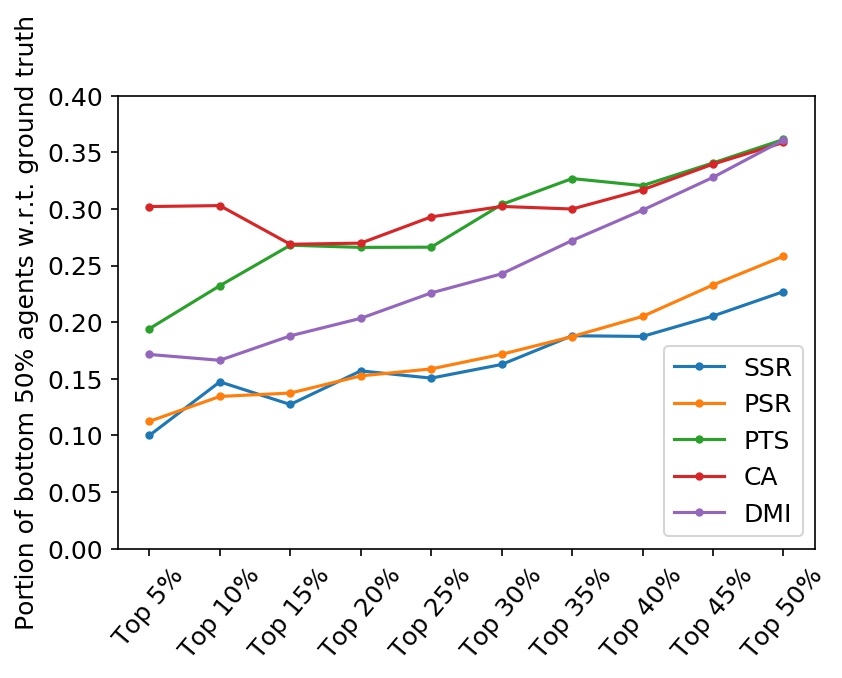}
        \caption{Mean squared loss}
    \end{subfigure}\hspace{1em}
    \begin{subfigure}[t]{0.3\textwidth}
        \centering
        \includegraphics[width=\textwidth]{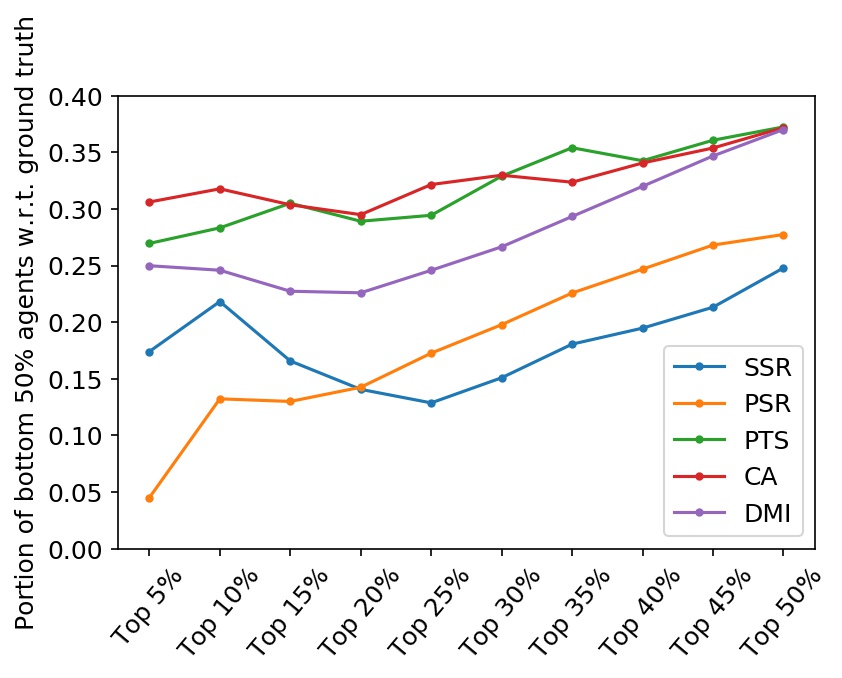}
        \caption{Mean cross-entropy loss}
    \end{subfigure}\hspace{1em}
    \begin{subfigure}[t]{0.3\textwidth}
        \centering
        \includegraphics[width=\textwidth]{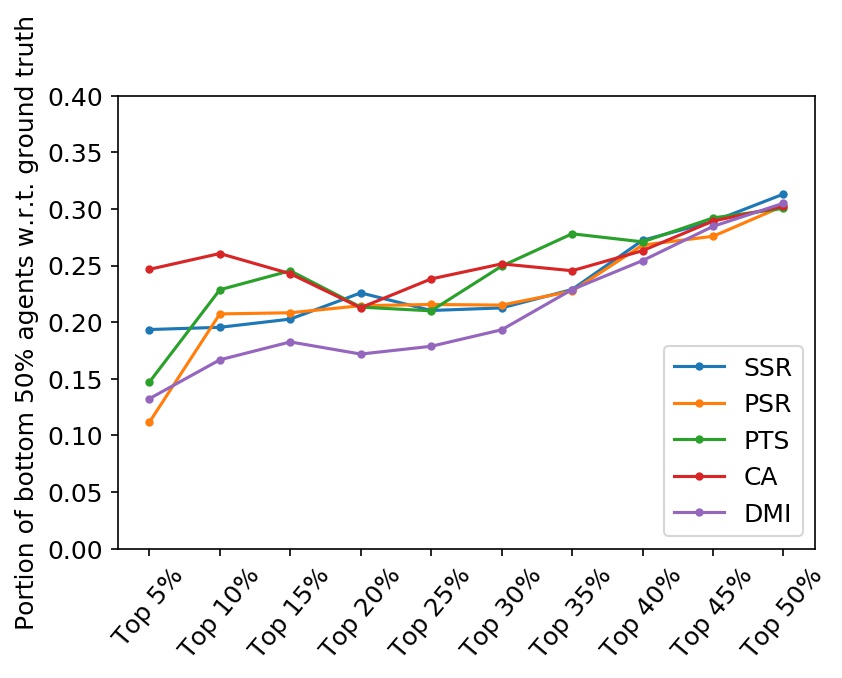}
        \caption{AUC-ROC}
    \end{subfigure}
    \vspace{-0.7em}
    \caption{The portion of bottom 50\% forecasters w.r.t. 3 different metrics (mean squared loss, cross-entropy loss, AUC-ROC loss) in the top $t$\% users selected by different methods (averaged over 14 datasets).\label{fig_bottom_user}}
    \vspace{-1em}
\end{figure*}

\paragraph{\textbf{\emph{Expert identification}}}
We exam to what extent different peer prediction scores can identify top performing experts. We rank the forecasters according to one of three most-widely used loss function (mean squared loss, mean cross-entropy loss, and AUC-ROC). We focus on two metrics about expert identification: i. percent of true top $t$\% forecasters in the top $t$\% forecasters selected by a peer prediction methods, ii. percent of below-average forecasters, the bottom 50\% forecasters, in the top $t$\% forecasters selected by a peer prediction methods. Results are shown in Fig~\ref{fig_top_user} and Fig~\ref{fig_bottom_user}. We find that for both mean squared loss and mean cross-entropy loss, in the top $t$\% forecaster selected by SSR, there are more true top $t$\% forecasters, than in the top forecasters selected by other peer prediction scores for $t$\% ranges from 5\% to 50\%. Meanwhile, there are less below-average forecasters in the top $t$\% forecasters top $t$\% by SSR and PSR than by the other peer prediction scores. For AUC-ROC, different peer prediction scores have similar performance, while SSR and DMI are slightly better than the others. These results echo the results about the correlation of peer prediction scores w.r.t. different SPSR.

\section{Concluding remarks}\label{sec:con}

We propose SSR to quantify the value of elicited information in IEWV settings, as strictly proper scoring rules do for the \emph{with} verification setting. SSR also induce truthful reporting in strictly dominant strategy for eliciting probabilistic predictions.  SSR contribute to both the SPSR and peer prediction literature. Our findings are both verified analytically and empirically. Our work opens up the study of calibrating the value of information for the peer prediction setting.

\begin{acks}
This research is based upon work supported in part by  National Science Foundation (NSF) under Grant No. CCF-1718549, the Office of the Director of National Intelligence (ODNI), Intelligence Advanced Research Projects Activity (IARPA), via 2017-17061500006 and the Defense Advanced Research Projects Agency (DARPA) and Space and Naval Warfare Systems Center Pacific (SSC Pacific) under Contract No. N66001-19-C-4014. The views and conclusions contained herein are those of the authors and should not be interpreted as necessarily representing the official policies, either expressed or implied, of NSF, ODNI, IARPA, DARPA, SSC Pacific or the U.S. Government. The U.S. Government is authorized to reproduce and distribute reprints for governmental purposes notwithstanding any copyright annotation therein.
\end{acks}

\bibliographystyle{ACM-Reference-Format}
\bibliography{noise_learning,library,myref}

\clearpage
\appendix
\appendix
\section*{{\LARGE Appendix}}
We fill in the missing proofs, the multi-outcome task extension and the experimental details.

\section{Missing Proofs}
\subsection{Proof of Lemma \ref{STOC:RELEVANT}}
\begin{proof}
Suppose not we will have 
\begin{align}
\Pr[y=0|z=0] = \Pr[y=0|z=1], \label{eqn:11}\\
\Pr[y=1|z=0] = \Pr[y=1|z=1].\label{eqn:22}
\end{align}
From Eqn. (\ref{eqn:11}) we know that
\begin{align*}
&\frac{\Pr[y=0,z=1]}{\Pr[z=1]} = 
\frac{\Pr[y=0,z=0]}{\Pr[z=0]} \\
\Leftrightarrow& \frac{\Pr[y=0]\nege{z}}{\Pr[z=1]} = \frac{\Pr[y=0](1-\nege{z})}{\Pr[z=0]},
\end{align*}
when $\Pr[y=0] \neq 0$ we know that
$$
\frac{\Pr[z=1]}{\Pr[z=0]} = \frac{\nege{z}}{1-\nege{z}}.
$$
Similarly from Eqn. (\ref{eqn:22}) we know 
$$
\frac{\Pr[z=1]}{\Pr[z=0]} = \frac{1-\pose{z}}{\pose{z}}.
$$
Therefore we obtained 
$$
\frac{\nege{z}}{1-\nege{z}} =  \frac{1-\pose{z}}{\pose{z}},
$$ 
from which we have $\nege{z}+\pose{z} = 1$. Contradiction.
\end{proof}

\subsection{Proof of Theorem \ref{noise:learn}}
\begin{proof}
Lemma 1, \citep{natarajan2013learning}, stated the version where $\pose{z}+\nege{z}<1$. Here we give the complete version where $\pose{z}+\nege{z}\ne1$. When $y=1$, we  have 
\begin{align*}
\E_{z|y=1}[R(q_i,z)] & 
= (1-\pose{z})R(q_i, 1) + \pose{z}R(q_i, 0)\\
&= (1-\pose{z})\frac{(1-\nege{z})S(q_i, 1)-\pose{z}S(q_i,0)}{1-\pose{z}-\nege{z}}+\pose{z}\frac{(1-\pose{z})S(q_i, 0)-\nege{z}S(q_i,1)}{1-\pose{z}-\nege{z}}\\
&=\frac{\left((1-\pose{z})(1-\nege{z})-\pose{z}\nege{z}\right)S(q_i, 1)}{1-\pose{z}-\nege{z}}\\
&=S(q_i, 1)
\end{align*}
When $y=0$, we have similarly
\begin{align*}
\E_{z|y=0}[R(q_i,z)] & 
= \nege{z} R(q_i, 1) + (1-\nege{z})R(q_i, 0)\\
&= \nege{z}\frac{(1-\nege{z})S(q_i, 1)-\pose{z}S(q_i,0)}{1-\pose{z}-\nege{z}}+(1-\nege{z})\frac{(1-\pose{z})S(q_i, 0)-\nege{z}S(q_i,1)}{1-\pose{z}-\nege{z}}\\
&=S(q_i, 0)
\end{align*}

\end{proof}

\subsection{Proof of Theorem \ref{THM:MAIN}}

\begin{proof}
The proof is straightforward following the ``unbiasedness'' property established for $\Sc(\cdot)$ in Lemma \ref{noise:learn}: let $y$ be a binary random variable drawn from any Bernoulli distribution, we have
\begin{align*}
\mathbb E_z[\Sc( q_i, z)] = \mathbb E_y\biggl[ \mathbb E_z[\Sc(q_i,z)|y]\biggr]  =  \mathbb E_y[S(q_i,y)].
\end{align*}
The theorem follows immediately from the strictly properness of $S$.

\end{proof}

\subsection{Proof of Theorem~\ref{THM:VAR}}

\begin{proof}

\begin{align*}
    &\mathbb E_{z}\bigl[ \Sc(q_i, z) - \mathbb E_{z}[\Sc(q_i, z) ] \bigr]^2 \\
    =& \frac{1}{(1-\pose{z}-\nege{z})^2} \cdot \biggl( \posz{z}\left(\Sc(q_i,1)-(\posz{z}\Sc(q_i,1)+(1-\posz{z})\Sc(q_i,0))\right)^2 \\
    &+(1-\posz{z})\left(\Sc(q_i,0)-(\posz{z}\Sc(q_i,1)+(1-\posz{z})\Sc(q_i,0))\right)^2\biggr)\\
    = &\frac{1}{(1-\pose{z}-\nege{z})^2} \cdot\biggl(\posz{z}(1-\posz{z})^2\left(\Sc(q_i,1) - \Sc(q_i,0) \right)^2\\
    &+(\posz{z})^2(1-\posz{z})\left(\Sc(q_i,1) - \Sc(q_i,0) \right)^2\biggr)\\
    =&\frac{\posz{z}(1-\posz{z})}{(1-\pose{z}-\nege{z})^2}\left(\Sc(q_i,1) - \Sc(q_i,0) \right)^2
\end{align*}
\end{proof}

\subsection{Proof of Lemma~\ref{lem_ref_cond} and Lemma~\ref{lemma_signal}} 
\begin{proof}

By Proposition~\ref{prop_posterior} and Assumption~\ref{ass_strategy} directly imply that 1) for each task, $q_{1,k},...,q_{N,k}$ are mutually independent conditioned on the ground truth $y_k$, 2) $(q_{1,k},...,q_{N,k},y_k)$ are i.i.d across tasks $k\in[M]$. 
As $z_{i,k}$ is independently drawn from Bernoulli$(\bar{q}_{-i,k})$, we immediately have that 1) for each task $k\in[M]$, $z_{i,k}$ is independent to $c_{i,k}$ and thus to $p_{i,k}:=\Pr[y_k=1|c_{i,k}]$, and 2) $(z_{i,k},y_k),k\in[M]$ are i.i.d. As a result of 2), $z_{i,k},k\in[M]$ have the same error rates w.r.t. the corresponding $y_k$. 
\end{proof}

\subsection{Proof of Theorem \ref{THM:UNIQUE2}}\label{sec:unique2}
\begin{proof}
Let $x^{-}:= \nege{z},~x^{+}:=1-\pose{z}.$ Recall the three equations we have
\begin{align}
&(1-p) \cdot x^{-} + p \cdot x^+ = \posz{-i}\label{eqn1}\\
&(1-p) \cdot (x^{-})^2 + p \cdot (x^+)^2 = \mat{-i}\label{eqn2}\\
&(1-p) \cdot (x^{-})^3 + p \cdot (x^+)^3 = \mathree{-i}\label{eqn3}
\end{align}

From Eqn.(\ref{eqn1}) we have
\[
(1-p)(x^--x^+) = \posz{-i} - x^+ \label{x-y}~~ (*).
\] 
Denote it as Eqn. $(*)$.
From Eqn.(\ref{eqn2}) and (\ref{eqn3}) we have
\begin{align*}
&(1-p)(x^--x^+)(x^-+x^+) + (x^+)^2 = \mat{-i}\\
&(1-p)(x^--x^+)\left((x^-)^2+x^- \cdot x^++ (x^+)^2\right) + (x^+)^3 = \mathree{-i}
\end{align*}
Plug Eqn.$(*)$ into above two equations we know, respectively:
\begin{align}
&(\posz{-i} - x^+)\cdot (x^-+x^+) + (x^+)^2 = \mat{-i} \nonumber \\ 
\Rightarrow& \posz{-i}(x^-+x^+)  - x^-\cdot x^+ = \mat{-i} \label{eqnc2}
\end{align}
\begin{align}
&(\posz{-i} - x^+)\left((x^-)^2+x^-\cdot x^++(x^+)^2\right) + (x^+)^3 = \mathree{-i}\nonumber \\
\Rightarrow & \posz{-i}\left((x^-)^2+x^-\cdot x^++(x^+)^2\right)  - x^-\cdot x^+(x^-+x^+) = \mathree{-i}\nonumber \\
\Rightarrow & \posz{-i}\left((x^-+x^+)^2 - x^-\cdot x^+\right) - x^-\cdot x^+(x^-+x^+) = \mathree{-i} \label{eqnc3}
\end{align}
Denote \[
x^-+x^+ = a, ~x^-\cdot x^+ = b
\]
then 
$
a  = \frac{b+\mat{-i}}{\posz{-i}}
$ from Eqn. (\ref{eqnc2}). Note the above is well defined, as o.w. if $\posz{-i} = 0$, we have to have $x^- = x^+ = 0$ which leads to $\nege{z}+\pose{z} = 1$, a contradiction. 

Substitute $(a,b)$ into Eqn. (\ref{eqnc3}) we have
\begin{align}
& \posz{-i}\cdot \left( \frac{(b+\mat{-i})^2}{(\posz{-i})^2}-b\right) - b\cdot \frac{b+\mat{-i}}{\posz{-i}} = \mathree{-i}\\
\Rightarrow & \frac{(b+\mat{-i})^2}{\posz{-i}} - b\cdot  \posz{-i} -\frac{b^2}{\posz{-i}} - \frac{b\cdot \mat{-i}}{\posz{-i}} = \mathree{-i}\\
\Rightarrow & \left( \frac{\mat{-i}}{\posz{-i}} - \posz{-i}\right) b = \mathree{-i} - \frac{(\mat{-i})^2}{\posz{-i}} \Rightarrow b = \frac{\posz{-i} \mathree{-i} - (\mat{-i})^2}{\mat{-i} - (\posz{-i})^2}
\end{align}
Further 
$
a = \frac{b+\mat{-i}}{\posz{-i}} = \frac{\mathree{-i}-\posz{-i}\mat{-i}}{\mat{-i} - (\posz{-i})^2}
$.
Now we show that $\mat{-i} \neq (\posz{-i})^2$, so the above pair of solutions are well defined. Suppose not, we will have
\begin{align}
&\left((1-p) \cdot x^- + p \cdot x^+\right)^2 = (1-p) \cdot (x^-)^2 + p \cdot (x^+)^2\\
\Rightarrow & 2\cdot p(1-p) \cdot x^-\cdot x^+ = p(1-p)\cdot (x^-)^2 + p(1-p) (x^+)^2\\
\Rightarrow & 2x^-\cdot x^+ = (x^-)^2 + (x^+)^2 \Rightarrow (x^--x^+)^2= 0 
\end{align}
which contradicts that $\nege{z}+\pose{z} \neq 1$. 

Then from
$x^-+x^+ = a, ~x^-\cdot x^+ = b,
$ we have
\begin{align}
x^{-} = \frac{a \pm \sqrt{a^2-4b}}{2}, x^+ = \frac{a \mp \sqrt{a^2-4b}}{2}, p= \frac{x^{-}-\posz{-i}}{x^{-}-x^{+}} \label{eqn:est} 
\end{align}

Denote the two pairs of solutions as $(x^{-}_1,x^+_1)$ and $(x^{-}_2,x^+_2)$. By symmetry we know 
$
x^{-}_1 = x^+_2,~x^+_1 = x^{-}_2. 
$ 
Further we have $x^{-}_1 \neq x^+_1$ and $x^{-}_2 \neq x^+_2$, due to the fact again that $\nege{z}+\pose{z} \neq 1$. Since $\posz{-i} = (1-p) \cdot x^{-} + p \cdot x^+$, and $p>0$, then it must be $\min\{x^-,x^+\} < \posz{-i} < \max\{x^-,y^+\}$. 

The above also implies that: If $p < 0.5$, we have $|x^+- \posz{-i}| \geq |\posz{-i} - x^{-}|$; else if $p > 0.5$,$|x^+ - \posz{-i}| \leq  |\posz{-i} - x^{-}|$. But we will note that if  $|x^+ - \posz{-i}| = |\posz{-i} - x^{-}|$, we can easily derive that $p=0.5$, contradicts with the assumption $p\ne0.5$.

\end{proof}

\subsection{Proof of Theorem \ref{THM:SUFFICIENT}}
\begin{proof}
We follow the shorthand notations as in the proof of Theorem \ref{THM:UNIQUE2} (Section \ref{sec:unique2}). Consider the fourth equation:
\begin{align}
(1-p)(x^-)^4 + p (x^+)^4 =& \left((1-p)(x^-)^3 + p (x^+)^3\right)(x^-+x^+) \\
&- x^-\cdot x^+\left ((1-p)(x^-)^2+p(x^+)^2\right).
\end{align}
$(1-p)(x^-)^3 + p (x^+)^3$ and $(1-p)(x^-)^2+p(x^+)^2$ are the second and third equation, while we know the first three equations already uniquely characterize $x^-+x^+$ and $x^-\cdot x^+$, so the fourth equation is redundant. This sets up the induction basis. For any $n>5$, we have 
\begin{align}
(1-p)(x^-)^n + p (x^+)^n &= \left ((1-p)(x^-)^{n-1} + p (x^+)^{n-1}\right)(x^-+x^+) \\
&- x^-\cdot x^+\left ((1-p)(x^-)^{n-2}+p (x^+)^{n-2}\right ). 
\end{align}
By induction hypothesis we know $(1-p)(x^-)^{n-1} + p (x^+)^{n-1}$ and $(1-p)(x^-)^{n-2}+p (x^+)^{n-2}$ can both be written as functions of the first three equations, so is $(1-p)(x^-)^n + p (x^+)^n$. Proved.
\end{proof}

\subsection{Proof of Theorem \ref{THM:ASYM}}

\begin{proof} 
With infinite number of tasks and agents, following Theorem~\ref{THM:UNIQUE2}, we can get the exact error rates $\pose{z},\nege{z}$ for the reference report $z_{i,k},k\in[M]$ we construct for agent $i$.  First for the case that ${\pose{z}} +{\nege{z}} = 1$,  it is indifferent for agent $i$ to truthfully report, or to misreport, or to randomize between the two strategies. Thus truth-telling is a weakly dominant strategy.  When ${\pose{z}}+{\nege{z}} \neq 1$, the dominant strategy argument follows from the strictly properness of (SSR\_alpha).

\end{proof}

\subsection{Proof of Lemma~\ref{lem_err}}

\begin{proof}
We consider the estimation of the error rates $\pose{z},\nege{z}$ of an agent $i$, and we consider a generic task as tasks are a priori similar. This, in the proof, we drop the subscript $k$, which indexes the tasks. There are two layers of estimation error is solving the system of equations Eqn. (\ref{eqn:1}, \ref{eqn:2}, \ref{eqn:3}):
\begin{itemize}
    \item \textbf{1. Estimation error due to heterogeneous agents:} the higher order equations doesn't capture the true matching probability with heterogeneous agents. As we draw $\zb$ and $\zc$ in a task without replacement, with finite number of agents, $\zb$ and $\zc$ are dependent with $\za$, and the error rates of $\zb$ and $\zc$ are not exactly the same to the error rates of $\za$ ($z$).
    \item \textbf{2. Estimation errors due to finite estimation samples:} The last sources of errors come from the estimation errors of $\widetilde{\mat{-i}}, \widetilde{\mathree{-i}}$ and $ \widetilde{\posz{-i}}$.
\end{itemize}

Next we bound the two errors separately.


\textbf{1. Estimation error due to heterogeneous agents:} The challenge lies in the fact that the higher order equations doesn't capture the true matching probability with heterogeneous agents. 

We first consider Eqn. (\ref{eqn:2}).
(\ref{eqn:2}) is not precise-- randomly picking a prediction signal from all agents without replacement leads to a different error rates. This will complicate the solution for the system of equations. We show that our estimation, though being ignoring the above bias, will not affect our results by too much: Let $k_1$ be the agent whose prediction signal is picked to be $\za$. Conditioned on agent $k_1$ being picked and on reports $q_1,...,q_N$, we have $\Pr[\za=\zb=1|q_1,...,q_N,k_1]=q_{k_1}\cdot\left(\frac{\sum_{j\ne i,{k_1}}q_{j}}{N-2}\right).$ 
Recall that $q_{k_1}$ is a random variable because of the private signal $c_{k_1}$ received by agent $k_1$ and the randomness in $\sigma_{k_1}$, and that $\pose{z}=\E_{q_1,...,q_N|y=1}[\bar{q}_{-i}]$. We have that
\begin{align*}
\Pr[\za=\zb=1|y=1]&=\E_{k_1}[\E_{q_1,...,q_N|y=1}[\Pr[\za=\zb=1|k_1,q_1,...,q_N]]]\\
& = \E_{k_1}\left[\E_{q_1,...,q_N|y=1}\left[q_{k_1}\cdot\left(\frac{\sum_{j\ne i,{k_1}}q_{j}}{N-2}\right)\right]\right]\\
& = \E_{k_1}\left[\E_{q_1,...,q_N|y=1}[q_{k_1}]\cdot\E_{q_1,...,q_N|y=1}\left[\frac{\sum_{j\ne i,{k_1}}q_{j}}{N-2}\right]\right]\\
& = \E_{k_1}\left[\E_{q_1,...,q_N|y=1}[q_{k_1}]\cdot\E_{q_1,...,q_N|y=1}\left[\frac{(N-1)\bar{q}_{-i}}{N-2}-\frac{q_{k_1}}{N-2}\right]\right]\\
& = \E_{k_1}\left[\E_{q_1,...,q_N|y=1}[q_{k_1}]\cdot\left(\frac{N-1}{N-2}\pose{z}-\frac{1}{N-2}\E_{q_1,...,q_N|y=1}[q_{k_1}]\right)\right]\\
& = \frac{N-1}{N-2}\pose{z}\E_{k_1}\left[\E_{q_1,...,q_N|y=1}[q_{k_1}]\right] - \frac{1}{N-2}\E_{k_1}\left[\E_{q_1,...,q_N|y=1}^2[q_{k_1}]\right]\\
& = \frac{N-1}{N-2}(\pose{z})^2 - \frac{1}{N-2}\E_{k_1}\left[\E_{q_1,...,q_N|y=1}^2[q_{k_1}]\right]\\
& = \frac{N-1}{N-2}(\pose{z})^2 - \frac{1}{N-2}\omega,
\end{align*}  
where $\omega:=\E_{k_1}\left[\E_{q_1,...,q_N|y=1}^2[q_{k_1}]\right]$.

Note both $\pose{z}$ and $\omega$ are no more than 1. Then ,
\begin{align*}
\left|\frac{N-1}{N-2}(\pose{z})^2 - \frac{1}{N-2}\omega - \frac{N-1}{N-2}(\pose{z})^2 \right|\le   \frac{(\pose{z})^2}{N-2}+\frac{1}{N-2}\omega \le \frac{2}{N-2}
\end{align*}
This adds $\frac{2}{N-2}$ error bias in the step where we replace $\Pr[z_1=z_2=1|y=1]$ with $(\pose{z})^2$ in the deduction of Eqn.~(\ref{eqn:2}). And, it finally adds $\frac{2}{N-2}$ error bias in estimating $\mat{-i}$ (through both $(\nege{z})^2$ and $(1-\pose{z})^2$) in Eqn.~(\ref{eqn:2}).

Similarly for the matching among three agents (Eqn.~(\ref{eqn:3})) we have
\[
\left|\Pr[\za=\zb=\zc=1|y=1] -(\pose{z})^3 \right| \leq \frac{3}{N-3}.
\]
And this adds $\frac{3}{N-3}$ error bias in estimating $\mathree{-i}$.



\textbf{2. Estimation errors due to finite estimation samples:} The last sources of errors come from the estimation errors of $\widetilde{\mat{-i}}, \widetilde{\mathree{-i}}$ and $ \widetilde{\posz{-i}}$. Direct application of the Chernoff bound gives us the following lemma:

\begin{lemma}\label{lemma:est:sample}
When there are $M$ samples for estimating $\widetilde{\mat{-i}}, \widetilde{\mathree{-i}}$ and $ \widetilde{\posz{-i}}$ respectively (total budgeting $3M$), we have with probability at least $1-\delta$ 
\[
|\widetilde{\mat{-i}}-\mat{-i}| \leq \sqrt{\frac{\ln \frac{6}{\delta}}{2M}}, ~|\widetilde{\mathree{-i}}-\mathree{-i}| \leq \sqrt{\frac{\ln \frac{6}{\delta}}{2M}}, ~|\widetilde{\posz{-i}}-\posz{-i}| \leq \sqrt{\frac{\ln \frac{6}{\delta}}{2M}}.
\]
\end{lemma}
The error analysis in \textbf{2} and \textbf{3} jointly imply that with probability at least $1-\delta$ 
\[
|\widetilde{\mat{-i}}-\mat{-i}| \leq \sqrt{\frac{\ln \frac{6}{\delta}}{2M}}+\frac{2}{N-2}, ~|\widetilde{\mathree{-i}}-\mathree{-i}| \leq \sqrt{\frac{\ln \frac{6}{\delta}}{2M}}+\frac{3}{N-3}, ~|\widetilde{\posz{-i}}-\posz{-i}| \leq \sqrt{\frac{\ln \frac{6}{\delta}}{2M}}.
\]

~\\

We now bound the error in estimating $\nege{z},\pose{z}$, under the $(\epsilon,\delta)$-event proved in Lemma \ref{lemma:est:sample}. 

We first introduce the following Lemma (Lemma 7 of \citep{sig15}): 
\begin{lemma}
For $k \geq 1$ and two sequences $\{l_i\}_{i=1}^k$ and $\{q_i\}_{i=1}^k$ and 
$
0 \leq l_i, q_i \leq 1, \forall i=1,...,k.
$, we have
\begin{align}
\left |\prod_{i=1}^k l_i - \prod_{j= 1}^k q_j\right | \leq \sum_{i=1}^k |l_i - q_i|~. \label{decomp1}
\end{align} 
\label{lemma:sep}
\end{lemma}

As a corrolary we prove the following:
\begin{corollary}
\label{coro:decomp}
For $k \geq 1$ and two pairs of numbers $(l_1,q_1), (l_2,q_2)$ and 
$
0 < l_i, q_i \leq 1, \forall i=1,2.
$, we have
\begin{align}
\left |\frac{l_1}{q_1} - \frac{l_2}{q_2}\right | \leq \frac{\sum_{i=1}^2 |l_i - q_i|}{q_1 q_2}~. \label{decomp2}
\end{align} 
\end{corollary}
\begin{proof}
This is because 
\begin{align*}
    &\left |\frac{l_1}{q_1} - \frac{l_2}{q_2}\right |\\
    =&\left |\frac{l_1 q_2 - l_2q_1}{q_1q_2}\right |\\
    \leq & \frac{\sum_{i=1}^2 |l_i - q_i|}{q_1q_2},
\end{align*}
where the last inequality we have used Lemma \ref{lemma:sep}.
\end{proof}
First of all, from Eqn. (\ref{eqn:est}) we can easily derive that 
\begin{align}
|\widetilde{\nege{z}}-\nege{z}| \leq \frac{|\tilde{a}-a|}{2}+\frac{|\sqrt{\tilde{a}^2-4\tilde{b}} - \sqrt{a^2-4b}|}{2}\\
    |\widetilde{\pose{z}}-\pose{z}| \leq \frac{|\tilde{a}-a|}{2}+\frac{|\sqrt{\tilde{a}^2-4\tilde{b}} - \sqrt{a^2-4b}|}{2}
\end{align}

\begin{align*}
    &\frac{|\sqrt{\tilde{a}^2-4\tilde{b}} - \sqrt{a^2-4b}|}{2}\\
    =&\frac{|(\sqrt{\tilde{a}^2-4\tilde{b}} - \sqrt{a^2-4b})\cdot (\sqrt{\tilde{a}^2-4\tilde{b}} + \sqrt{a^2-4b})|}{2(\sqrt{\tilde{a}^2-4\tilde{b}} + \sqrt{a^2-4b})}\\
    \leq &\frac{\tilde{a}^2-a^2}{2 \sqrt{a^2-4b}} +\frac{4|\tilde{b}-b|}{2 \sqrt{a^2-4b}} \\
      \le &\frac{|\tilde{a}-a|^2}{\sqrt{a^2-4b}} +\frac{a \cdot |\tilde{a}-a|}{\sqrt{a^2-4b}}+\frac{2|\tilde{b}-b|}{ \sqrt{a^2-4b}} \\
\end{align*}

To summarize 
\begin{align}
|\widetilde{\nege{z}}-\nege{z}| \leq \frac{1}{2}\biggl( (1+\frac{a}{\sqrt{a^2-4b}})|\tilde{a}-a| +\frac{2|\tilde{b}-b|}{ \sqrt{a^2-4b}} + \frac{1}{\sqrt{a^2-4b}} |\tilde{a}-a|^2 \biggr)\\
    |\widetilde{\pose{z}}-\pose{z}| \leq \frac{1}{2}\biggl( (1+\frac{a}{\sqrt{a^2-4b}})|\tilde{a}-a| +\frac{2|\tilde{b}-b|}{ \sqrt{a^2-4b}} + \frac{1}{\sqrt{a^2-4b}} |\tilde{a}-a|^2 \biggr)
\end{align}

The key tasks here reduce to bounding $|\tilde{a}-a|$ and $|\tilde{b}-b|$. Recall 
\begin{align*}
a:=&     \frac{\mathree{-i}-\posz{-i}\mat{-i}}{\mat{-i} - (\posz{-i})^2}\\
b:=& \frac{\posz{-i} \mathree{-i} - (\mat{-i})^2}{\mat{-i} - (\posz{-i})^2}
\end{align*}

We know the following facts 
\begin{align*}
&| (\widetilde{\mat{-i}}-(\widetilde{\posz{-i}})^2) - (\mat{-i}-(\posz{-i})^2) | \\
\leq & |(\widetilde{\posz{-i}})^2-(\posz{-i})^2| + |\widetilde{\mat{-i}}-\mat{-i}| \\
\leq & 2|\widetilde{\posz{-i}}-\posz{-i}| + |\widetilde{\mat{-i}}-\mat{-i}|\\
\leq & 3\sqrt{\frac{\ln \frac{6}{\delta}}{2M}} + \frac{2}{N-2},
\end{align*}
\begin{align*}
&|(\widetilde{\mathree{-i}}-\widetilde{\mat{-i}} \widetilde{\posz{-i}})-(\mathree{-i}-\mat{-i} \posz{-i})| \\
\leq & |\widetilde{\mathree{-i}}-\mathree{-i}|+|\widetilde{\mat{-i}} \widetilde{\posz{-i}}-\mat{-i} \posz{-i}|\\
\leq &  |\widetilde{\mathree{-i}}-\mathree{-i}|+|\widetilde{\mat{-i}} -\mat{-i}|+|\widetilde{\posz{-i}}- \posz{-i}|\\
 \leq & 3\sqrt{\frac{\ln \frac{6}{\delta}}{2M}} + \frac{2}{N-2}+\frac{3}{N-3},
 \end{align*}
 
 \begin{align*}
     &|(\widetilde{\posz{-i}} \widetilde{\mathree{-i}} - (\widetilde{\mat{-i}})^2)-(\posz{-i} \mathree{-i} - (\mat{-i})^2)|\\
     \leq &|\widetilde{\posz{-i}}-\posz{-i}| +  |\widetilde{\mathree{-i}}-\mathree{-i}|+2|\widetilde{\mat{-i}}-\mat{-i}|\\
     \leq & 4\sqrt{\frac{\ln \frac{6}{\delta}}{2M}} + \frac{2}{N-2}+\frac{3}{N-3},
\end{align*}

Next we prove that 
\begin{align*}
 &\mat{-i} - (\posz{-i})^2 \\
 = &(1-p)\cdot (x^{-})^2 + p\cdot (x^{+})^2- ((1-p)\cdot x^{-} + p\cdot x^{+})^2\\
  =& (1-p)\cdot p\cdot  (x^{-}-x^{+})^2\\
  = & (1-p)\cdot p\cdot \frac{\Delta^2}{(1-p)^2}\\
  =&\frac{p\cdot \Delta^2}{1-p}
\end{align*}
where the third equality is due to
$\Delta:=(1-p) (1-\nege{z}-\pose{z})=(1-p)({x^-}-{x^+})$.

Let $N$ be sufficiently large such that
\[
3\sqrt{\frac{\ln \frac{6}{\delta}}{2M}} + \frac{2}{N-2} < \frac{p\cdot \Delta^2}{2(1-p)}
\]
then 
\[
\widetilde{\mat{-i}}-(\widetilde{\posz{-i}})^2 \geq \frac{p\cdot \Delta^2}{2(1-p)}
\]

Therefore 
\begin{align*}
|\tilde{a}-a| \leq& \frac{| (\widetilde{\mat{-i}}-(\widetilde{\posz{-i}})^2) - (\mat{-i}-(\posz{-i})^2) |  +|(\widetilde{\mathree{-i}}-\widetilde{\mat{-i}} \widetilde{\posz{-i}})-(\mathree{-i}-\mat{-i} \posz{-i})|  }{|\widetilde{\mat{-i}}-(\widetilde{\posz{-i}})^2|\cdot |\mat{-i} - (\posz{-i})^2 |}\\
\leq & \frac{2(1-p)^2}{p^2 \Delta^4}\bigl( 6\sqrt{\frac{\ln \frac{6}{\delta}}{2M}} + \frac{4}{N-2}+\frac{3}{N-3}\bigr)
\end{align*}
Similarly for $b$
\begin{align*}
|\tilde{b}-b| \leq& \frac{2(1-p)^2}{p^2 \Delta^4}\bigl( 7\sqrt{\frac{\ln \frac{6}{\delta}}{2M}} + \frac{4}{N-2}+\frac{3}{N-3}\bigr)
\end{align*}

Together we proved that when 
\[
3\sqrt{\frac{\ln \frac{6}{\delta}}{2M}} + \frac{2}{N-2} < \frac{p\cdot \Delta^2}{2(1-p)},
\]
we have
\begin{align*}
|\widetilde{\nege{z}}-\nege{z}|  \leq O\bigl(\sqrt{\frac{\ln \frac{1}{\delta}}{2M}}+\frac{1}{N}\bigr)\\
|\widetilde{\pose{z}}-\pose{z}|  \leq O\bigl(\sqrt{\frac{\ln \frac{1}{\delta}}{2M}}+\frac{1}{N}\bigr)
\end{align*}
\end{proof}

\subsection{Proof of Theorem~\ref{THM:EST}}
\begin{proof}


This proof is straight-forward following the error rate bounding result (Lemma~\ref{lem_err}):
Using $sgn(z),z\in\{0,1\}$ as the superscript, where $sgn(0)$ refers to super script ``$-$'' and $sgn(1)$ refers to super script ``$+$''. We have
\begin{align*}
    &|\widetilde{R}(p,z)-R(p,z)| \\
    =&\biggl| \left(\frac{1-\widetilde{\sgne{sgn(1-z)}{z}}}{1-\widetilde{\pose{z}}-\widetilde{\nege{z}}} - \frac{1-\sgne{sgn(1-z)}{z}}{1-\pose{z}-\nege{z}}\right)S(p, z) \\
&- \left(\frac{\widetilde{\sgne{sgn(z)}{z}}}{1-\widetilde{\pose{z}}-\widetilde{\nege{z}}} - \frac{\sgne{sgn(z)}{z}}{1-\pose{z}-\nege{z}}\right)S(p, 1-z)\biggr|\\
\leq &\biggl|\frac{1-\widetilde{\sgne{sgn(1-z)}{z}}}{1-\widetilde{\pose{z}}-\widetilde{\nege{z}}} - \frac{1-\sgne{sgn(1-z)}{z}}{1-\pose{z}-\nege{z}}\biggr| \max S \\
&- \biggl|\frac{\widetilde{\sgne{sgn(z)}{z}}}{1-\widetilde{\pose{z}}-\widetilde{\nege{z}}} - \frac{\sgne{sgn(z)}{z}}{1-\pose{z}-\nege{z}}\biggr| \max S
\end{align*}

Since $\epsilon \leq (1-\nege{z}-\pose{z})/4$ we know that 
\[
1-\widetilde{\pose{z}}-\widetilde{\nege{z}} > (1-\nege{z}-\pose{z})/2
\]
Using Corollary~\ref{coro:decomp} we know
\begin{align*}
    \biggl|\frac{1-\widetilde{\sgne{sgn(1-z)}{z}}}{1-\widetilde{\pose{z}}-\widetilde{\nege{z}}} - \frac{1-\sgne{sgn(1-z)}{z}}{1-\pose{z}-\nege{z}}\biggr| \leq \frac{3\epsilon}{\Delta^2/2} = \frac{6\epsilon}{\Delta^2}\\
    \biggl|\frac{\widetilde{\sgne{sgn(z)}{z}}}{1-\widetilde{\pose{z}}-\widetilde{\nege{z}}} - \frac{\sgne{sgn(z)}{z}}{1-\pose{z}-\nege{z}}\biggr| \leq \frac{3\epsilon}{\Delta^2/2} = \frac{6\epsilon}{\Delta^2}
\end{align*}
Plug back we have proved the claim:
\begin{align*}
   & |\E[\widetilde{R}(p,z)]-\E[R(p,z)]| \leq \E|\widetilde{R}(p,z)]-R(p,z)| \leq \frac{12\epsilon \cdot \max S}{\Delta^2},\\
    &|\E[\widetilde{R}(p,z)]-\E[S(p,y)]| = |\E[\widetilde{R}(p,z)]-\E[R(p,z)]|  \leq \frac{12\epsilon \cdot \max S}{\Delta^2}~.
\end{align*}

\end{proof}

\subsection{Proof of Theorem \ref{FINITE}}


\begin{proof}

W.l.o.g, we consider the case where after the estimation of the error rates, we have $\tilde{\pose{z}}+\tilde{\nege{z}}<1$. When $\tilde{\pose{z}}+\tilde{\nege{z}}>1$, the value given by SSR will be keep same if we do the following transformation: $z'\leftarrow 1-z;$ $\tilde{\pose{z}}' \leftarrow 1- \tilde{\pose{z}};$ $\tilde{\nege{z}}'\leftarrow1-\tilde{\nege{z}}.$ Then, we reduce the case where $\tilde{\pose{z}}+\tilde{\nege{z}}>1$ to case a where $\tilde{\pose{z}}+\tilde{\nege{z}}<1$.

We consider an arbitrary agent $i\in[N]$ and the analysis can be applied to each agent. As tasks are a-priori-identical to an agent, we also drop the subscript $k$, while keep in mind that we consider a single task. The theorem to prove talks about the incentive of the \SSRM. As mentioned, w.l.o.g., we can multiple a positive constant to the scoring rule without influence the incentive properties, if the constant does not depend on the reports of the agent who is being scored. Therefore, when scoring agent $i$, we multiple $1-\tilde{\pose{z}}-\tilde{\nege{z}}$ to the SSR for her. This helps eliminate the denominator of the SSR and will ease our analysis without influencing the correctness of the analysis.

Recall that $p_i$ is the posterior belief of agent $i$. When we take expectation w.r.t. $y$, $y$ is drawn from Bernoulli($p_i$). When we take expectation w.r.t. $z$, $z$ is generated by $y$ based on the true error rates $\pose{z},\nege{z}$ while $y\sim\text{Bernoulli}(p_u)$. When we use $sgn(z),z\in\{0,1\}$ as the superscript, $sgn(0)$ refers to super script ``$-$'' and $sgn(1)$ refers to super script ``$+$''.

Suppose that $S(q_{i},z)$ is strongly concave w.r.t. $q_{i}$ $\forall z\in\{0,1\}$ with parameter $\lambda$. We have
\begin{align*}
&\mathbb E[\widetilde{\Sc}(p_i,z)] - \mathbb E[\widetilde{\Sc}(q_i,z)]\\
=&\left(\mathbb E[\widetilde{\Sc}(p_i,z)] - \mathbb E[\Sc(p_i,z)] \right) -  \left(\mathbb E[\widetilde{\Sc}(q_i,z)] - \mathbb E[\Sc(q_i,z)] \right) +  \left (\mathbb E[\Sc(p_i,z)]-\mathbb E[\Sc(q_i,z)] \right)
\end{align*}

For the last term $\mathbb E[\Sc(p_i,z)]-\mathbb E[\Sc(q_i,z)]$, noticing that using strongly concavity we have
\begin{align*}
\mathbb E[\Sc(p_i,z)]-\mathbb E[\Sc(q_i,z)]= \mathbb E[S(p_i,y)] - \mathbb E[S(q_i,y)] \geq \lambda |p_i - q_i|.
\end{align*}

Now we analyze the first two terms $\mathbb E[\widetilde{\Sc}(p_i,z)] - \mathbb E[\Sc(p_i,z)]$ and $ \mathbb E[\widetilde{\Sc}(q_i,z)] - \mathbb E[\Sc(q_i,z)]$.
Let $\epsilon(x,z) := 
\widetilde{\Sc}(x,z) - \Sc(x,z)
$ be the error term for any report $x\in[0,1]$.
We notice that
\begin{align*}
&\epsilon(x,z) =\widetilde{\Sc}(x,z) - \Sc(x,z)\\
=&\left((1-\widetilde{\sgne{sgn(1-z)}{z}}) - (1-\sgne{sgn(1-z)}{z})\right)S(x, z) \\
&- \left((\widetilde{\sgne{sgn(z)}{z}}) - (\sgne{sgn(z)}{z})\right)S(x, 1-z).
\end{align*}
Due to the sample complexity results we know that with probability at least $1-\delta$ that
\begin{align*}
&\left |(1-\widetilde{\sgne{sgn(1-z)}{z}}) - (1-\sgne{sgn(1-z)}{z})\right | \leq \epsilon,\\
&\left |(\widetilde{\sgne{sgn(z)}{z}}) - (\sgne{sgn(z)}{z})\right | \leq \epsilon,
\end{align*}
where   $\epsilon = O\left(\frac{1}{N}+\sqrt{\frac{\log \frac{1}{\delta}}{M}}\right)$.

Suppose $S(x,z)$ is also Lipschitz w.r.t. $x$ $\forall z\in\{0,1\}$ with parameter $L$. By Lipschitz conditions we know that with probability at least $1-\delta$ that $\epsilon(x,z) $ is Lipschitz with parameters $2\epsilon L$ by composition property and thus 
\begin{align*}
|\epsilon(p_i,z) -  \epsilon(q_i,z)| \leq 2\epsilon L \cdot |p_i - q_i|. 
\end{align*}
Combining we have 
\begin{align*}
|\mathbb E[\epsilon(p_i,z)] -  \mathbb E[\epsilon(q_i,z)]| \leq 2\epsilon L \cdot |p_i - q_i| + 2\delta L\cdot |p_i - q_i|. 
\end{align*}
Therefore when $\epsilon,\delta$ are small enough such that 
\[
2\epsilon L + 2\delta L < \lambda, 
\]
no deviation is profitable. 

\end{proof}

\section{SSR for multi-outcome tasks}
\label{sec_non-binary}
In this section, we extend the surrogate scoring rule to tasks with more than two outcomes.  
Consider a multi-outcome task with $C+1$ possible outcomes (classes). We denote  the ground truth of task by $y\in\mathcal{C}$ and a noisy copy of the ground truth by $z\in\mathcal{C}$, where $\mathcal{C}=\{0,...,C\}$. 
$z$ has the following confusion matrix w.r.t. $y$:
\[ E_z = 
\begin{bmatrix}
    e_{0,0}       & e_{0,1} &  \dots & e_{0,C} \\
    e_{1,0}        & e_{1,1}  & \dots & e_{1, C} \\
    \hdotsfor{4} \\
    e_{C,0}       & e_{C,1} & \dots & e_{C,C}
\end{bmatrix}.
\]
Each entries $e_{u,v}$ indicates the flipping probability of $z$:
$e_{u,v} = \Pr[z = v|y=u].$ Let $\Delta^c:=\{(x_0,...,x_c)|\sum_{i=0}^c x_i=1,x_0,...,x_c\ge0\}$ be the $C$-dimensional probability simplex.
The surrogate scoring rules for a task with $C+1$ outcomes is defined as follows.
\begin{definition}[Surrogate Scoring Rules]
$\Sc: \Delta^C \times \mathcal{C}\rightarrow \mathbb R_+$ is a surrogate scoring rule for a $(C+1)$-outcome task if for some strictly proper scoring rule $S:  \Delta^C \times \mathcal{C} \rightarrow \mathbb R_+$ and a strictly increasing function $f:\mathbb R_+ \rightarrow \mathbb R_+$, the following equation holds:
$$\forall \mathbf{p},\mathbf{q}\in\Delta^C,\forall E_z\in[0,1]^{(C+1)\times(C+1)} (E_z\text{ is invertible}): \underline{\mathbb E_{z}[\Sc(\mathbf{q}, z)] = f\left( \mathbb E_y[S(\mathbf{q}, y)]\right)},$$ where the ground truth $y\sim\text{Categorical}(\mathbf{p})$ and $z$ is a noisy copy generated by $y$ according to the confusing matrix $E_z$.
\end{definition}

We have the following theorem immediately. 

\begin{theorem}
Given an agent's fixed prior $\mathbf{p}$ and private signal $c_i$, SSR $R(\mathbf{q}, z)$ with noisy ground truth $z$ is strictly proper for eliciting the agent's posterior $\mathbf{p}_i$ ($\Pr[y|c_i]$) if $z$ and $c_i$ are independent conditioned on $y$ and $E_z$ is invertible. 
\end{theorem}

Now we give an implementation SSR\_alpha for a multi-outcome task. Let $S(\mathbf{q}_i)$ be the vector of SPSR scores for $\mathbf{q}_i$ under each realizations of $y$, i.e., $S(\mathbf{q}_i):=(S(\mathbf{q}_i,y=0),..., S(\mathbf{q}_i,y=C))$. Similarly, let $R(\mathbf{q}_i):=(R(\mathbf{q}_i, z=0),..., R(\mathbf{q}_i, z=C) )$.  Our implementation SSR\_alpha goes as follows:
$$R(\mathbf{q}_i) := {(E_z)}^{-1}S(\mathbf{q}_i)$$

Clearly, for SSR\_alpha we have $S(\mathbf{q}_i) = E_zR(\mathbf{q}_i)$, which gives
$$\forall v\in\mathcal{C}, S(\mathbf{q}_i,y=v) =\sum_{k=0}^C e_{v,k}R(\mathbf{q}_i,z=k) = \E_{z|y=v}[R(\mathbf{q}_i, z)]. $$

\begin{lemma}
For \text{(SSR\_alpha)}:$\forall y\in\mathcal{C},
\mathbb E_{z|y}[\Sc(p_i, z) ] = 
S(p_i, y)$
\end{lemma}
The following theorem follows immediately.

\begin{theorem}
\text{(SSR\_alpha)} is a surrogate scoring rule for a multi-outcome task, and for any distribution $\mathbf{p}\in\Delta^C$ of the ground truth $y$ and for any invertible confusing matrix $E_z$ of the noisy signal $z$, we have  $\forall \mathbf{q}\in\Delta^C,  \E_{z}[\Sc(\mathbf{q}, z) ] = 
\mathbb E_y[S(\mathbf{q}, y)]$.
\end{theorem}

We include a detailed example of SSR\_alpha for three-outcome tasks below.

\begin{example}
Let C = 2, i.e., $\mathcal{C}=\{0,1,2\}$. Let the confusing matrix of a noisy signal $z$ being
\[ E_z = 
\begin{bmatrix}
    0.5      & 0.25 &  0.25 \\
    0.25       & 0.5  & 0.25 \\
    0.25       & 0.25  & 0.5 
\end{bmatrix} \Rightarrow {(E_z)}^{-1} = 
\begin{bmatrix}
   3     & -1 &  -1 \\
    -1       & 3  & -1 \\
    -1       & -1  & 3 
\end{bmatrix}
\]
We obtain a closed-form of SSR\_alpha:
\begin{align*}
R(\mathbf{q},z=0)&:= 3  S(\mathbf{q},0)  - S(\mathbf{q},1) - S(\mathbf{q},2) \\ 
R(\mathbf{q},z=1)&:= -  S(\mathbf{q},0)  +3 S(\mathbf{q},1) - S(\mathbf{q},2) \\ 
R(\mathbf{q},z=2)&:= -  S(\mathbf{q},0)  - S(\mathbf{q},1) + 3 S(\mathbf{q},2)
\end{align*}
\end{example}

\section{More on experiments}
\subsection{Datasets}

\paragraph{\textbf{\emph{GJP datasets~\cite{atanasov2016distilling}}}} It contains four datasets on geopolitical forecasting questions. The four datasets, denoted by G1$\sim$G4, was collected from 2011 to 2014 respectively. They have different forecasting questions and forecasters. Forecasters were all recruited from professional societies’ email lists, blogs, research centers, alumni associations, and personal connections. Each forecaster has a single probabilistic prediction for a question she answered in the datasets. 

\paragraph{\textbf{\emph{HFC datasets~\cite{HFC}}}} It contains three datasets, denoted by H1$\sim$H3, collected from the Hybrid Forecast Competition organized by IARPA in 2018. The three datasets share the same forecasting questions about geopolitics, finance, economics, etc, but have different forecasters and collecting methods. H1 was independently collected by the Hughes Research Laboratories (HRL) with forecasters recruited from Amazon Mechanical Turk (AMT). H2, H3 was collected by collected by IARPA. The forecasters in H2 were recruited from Amazon Mechanical Turk (AMT), while the forecasters in H3 were volunteers who knew the project by email, blog, etc and voluntarily took part in the project. These three datasets record multiple probabilistic predictions each forecaster made at different dates. We used the final prediction made by a forecaster on a question she answered.

\paragraph{\textbf{\emph{MIT datasets~\cite{prelec2017solution}}}} It contains seven datasets, denoted as M1a, M1b, M1c, M2, M3, M4a, M4b, with different questions and forecasters. The questions ranges from the capital of states to the price interval that artworks belong to, to some trivia questions. The forecasters were students in class and colleagues in labs. In datasets M1a, M1b, M4a, M4b, forecasters made binary vote on a forecasting question. In datasets M1c, M2, M3, forecasters gave a probabilistic prediction. 

\subsection{Correlation with SPSR (Spearman's test)}

\begin{figure*}[h]
   \centering
    \begin{subfigure}[t]{0.27\textwidth}
        \centering
        \includegraphics[width=\textwidth]{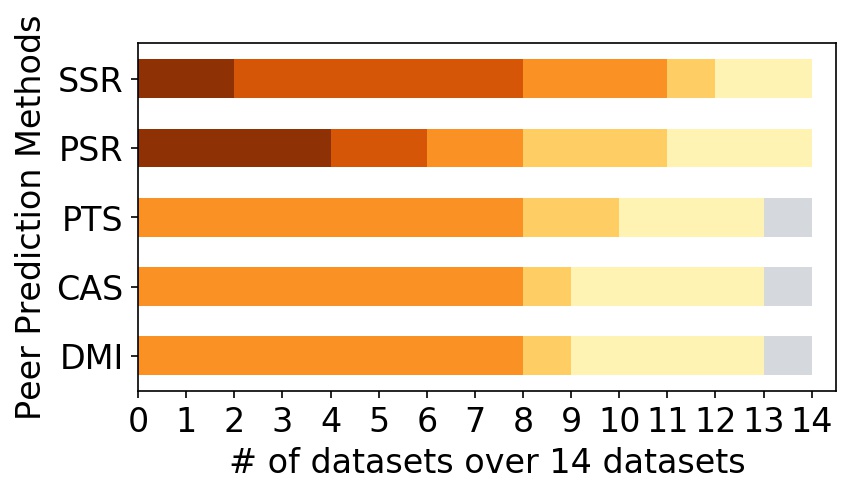}
        \caption{Brier score}
    \end{subfigure}  
    \begin{subfigure}[t]{0.27\textwidth}
        \centering
        \includegraphics[width=\textwidth]{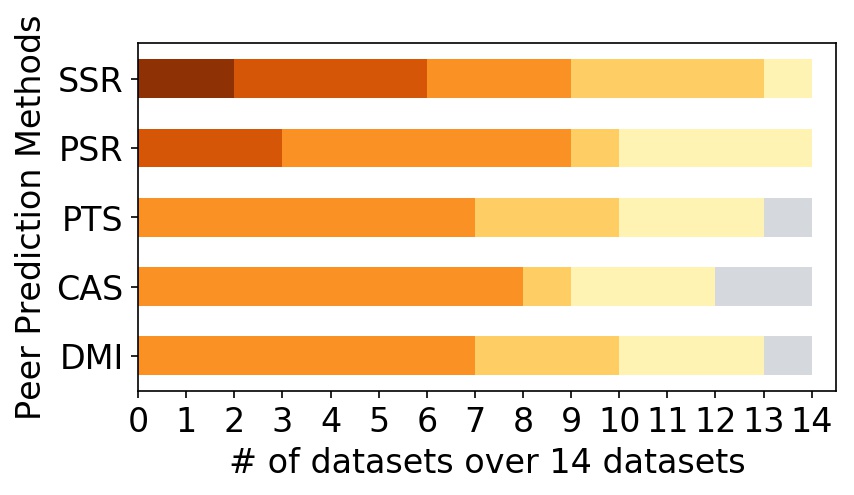}
        \caption{Log scoring rule}
    \end{subfigure}  
    \begin{subfigure}[t]{0.27\textwidth}
        \centering
        \includegraphics[width=\textwidth]{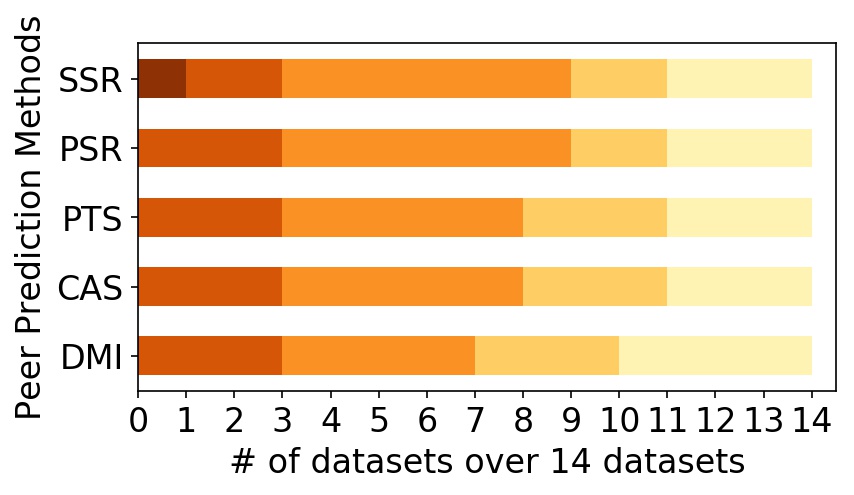}
        \caption{Rank sum scoring rule}
    \end{subfigure}
    \begin{subfigure}[t]{0.15\textwidth}
        \centering
        \includegraphics[width=\textwidth]{correlation_full_legend.jpg}
    \end{subfigure}
    \caption{The number of datasets in each level of correlation (measured by Spearman's correlation coefficient) between individuals' peer prediction scores and different SPRS\label{fig_corr_spearman}}
\end{figure*}

\end{document}